\documentclass[a4paper,fleqn,usenatbib]{mnras}

\usepackage[T1]{fontenc}
\usepackage{ae,aecompl}
\usepackage{amsmath}
\usepackage{amssymb}
\usepackage{graphicx}
\usepackage{hyperref}
\usepackage{subcaption}	
\usepackage{epstopdf}	
\usepackage{verbatim}	
\numberwithin{equation}{section}
\title[Sparse GPs for heteroscedastic noise]{{\sc GPz}: Non-stationary sparse Gaussian processes for heteroscedastic uncertainty estimation in photometric redshifts}
\author[Almosallam et al.]
{Ibrahim A. Almosallam,$^{1,2}$\thanks{E-mail: ialmosallam@kacst.edu.sa} Matt J. Jarvis$^{3,4}$ and Stephen J. Roberts$^{2}$\\
$^1$King Abdulaziz City for Science and Technology, Riyadh 1142, Saudi Arabia\\
$^2$Information Engineering, Parks Road, Oxford, OX1 3PJ, UK\\
$^3$Oxford Astrophysics, Department of Physics, Keble Road, Oxford, OX1 3RH, UK\\
$^4$Department of Physics, University of the Western Cape, Bellville 7535, South Africa}

\date{\today}

\pubyear{2016}

\begin{document}

\label{firstpage}
\pagerange{\pageref{firstpage}--\pageref{lastpage}}
\maketitle

\begin{abstract}
The next generation of cosmology experiments will be required to use photometric redshifts rather than spectroscopic redshifts. Obtaining accurate and well-characterized photometric redshift distributions is therefore critical for {\em Euclid}, the Large Synoptic Survey Telescope and the Square Kilometre Array. However, determining accurate variance predictions alongside single point estimates is crucial, as they can be used to optimize the sample of galaxies for the specific experiment (e.g. weak lensing, baryon acoustic oscillations, supernovae), trading off between completeness and reliability in the galaxy sample.
The various sources of uncertainty in measurements of the photometry and redshifts put a lower bound on the accuracy that any model can hope to achieve. The intrinsic uncertainty associated with estimates is often non-uniform and input-dependent, commonly known in statistics as \emph{heteroscedastic} noise.
However, existing approaches are susceptible to outliers and do not take into account variance induced by non-uniform data density and in most cases require manual tuning of many parameters. In this paper, we present a Bayesian machine learning approach that jointly optimizes the model with respect to both the predictive mean and variance we refer to as Gaussian processes for photometric redshifts ({\sc GPz}). The predictive variance of the model takes into account both the variance due to data density and photometric noise. Using the SDSS DR12 data, we show that our approach substantially outperforms other machine learning methods for photo-z estimation and their associated variance, such as {\sc tpz} and {\sc annz2}.
We provide a {\sc matlab} and {\sc python} implementations that are available to download at \url{https://github.com/OxfordML/GPz}.
\end{abstract}


\section{Introduction}
\label{sec-introduction}
Photometric redshift estimation largely falls into two main methodological classes, machine learning and template fitting. Machine learning methods such as artificial neural networks \citep[e.g. {\sc annz};][]{Firth2003,Collister04,Sadeh2015}, nearest-neighbour \citep[NN;][]{Ball2008}, genetic algorithms \citep[e.g.][]{Hogan2015}, self-organized maps \citep[][]{Geach2012,Masters2015}, random forest \citep[{\sc tpz};][]{kind2013} and Gaussian processes \citep[GPs;][]{Way2009,bonfield10,almosallam2015}, use different statistical models to predict the most probable redshift given the observed photometry, using a training sample where usually the spectroscopic redshift is known. Artificial neural networks motivate the most commonly used machine learning method \citep{Firth2003,vanzella2004photometric,brescia2014catalogue,Sadeh2015}. The parameters of the models often cannot be analytically inferred, so global and greedy optimization methods are used to estimate their parameters. In addition to providing a point estimate, machine learning methods can provide the degree of uncertainty in their predictions \citep{roberts1996,bishop2006,kind2013,bonnett2015,rau2015}. Template fitting methods on the other hand do not learn a model from a training sample but rather use templates of galaxy spectral energy distributions (SEDs) for different galaxy types that can be redshifted to fit the photometry. Some limitations of template fitting methods are whether the templates are representative of the galaxies observed at high redshift and how emission lines affect the photometry. Some allow spectroscopic data to be used to adjust the zero points on the photometry to compensate for any slight mismatch between SED templates and the observations.
Examples of template fitting software include {\sc hyperz}; \citep[][]{Hyperz}, {\sc zebra}; \citep{ZEBRA}, {\sc eazy}; \citep[][]{EAZY} and {\sc le phare} \citep[][]{Ilbert2006}. There have been comprehensive evaluations of different photometric redshift estimation techniques \citep{hildebrandt10,abdalla11,sanchez14,bonnett2015}. 

In this paper, we extend our previous work \citep{almosallam2015} and complete the Bayesian picture of the sparse Gaussian model. In \cite{almosallam2015} the noise variance was assumed to be constant and treated as an input parameter optimized using cross-validation. In the approach we propose here, the variance is an input-dependent function and is learned jointly with the mean function. The variance produced by the proposed approach is composed of two terms that captures different sources of uncertainty. The first term is the intrinsic uncertainty about the mean function due to data density, whereas the second term captures the uncertainty due to the intrinsic noise or the lack of precision/features in the training set. This provides additional utility to identify regions of input space where more data is required, versus areas where additional precision or information is required. Classical Gaussian processes (GPs), for example, only model the uncertainty about the mean function and assume that the noise uncertainty is constant.

Such a method is particularly useful for machine learning-based methods, as it is often the case that the training samples are incomplete due to the difficulty in obtaining complete spectroscopic redshift information. Moreover, imaging data is predominantly many magnitudes deeper than spectroscopic data, therefore quantifying the noise on the photometric redshift in terms of whether it is due to the density of training data available in a certain colour space, or due to a lack of sufficiently precise data within the colour space of interest could be crucial. In particular it means that optimal spectroscopic survey strategies can be implemented to increase the photometric accuracy in the best way possible for a given experiment, i.e. obtaining more data in different colour space, or improving the quality of data in that colour space through additional imaging for example.

This paper is organized as follows; first a summary of related work is presented in \autoref{sec-related-work} followed by an overview of sparse Gaussian processes in \autoref{sec-sparse-gaussian-processes}. In \autoref{sec-automatic-relevance-determination} we discuss how to expand the method to favour simpler, or sparser, models via automatic relevance determination. The extension to account for heteroscedastic noise is described in \autoref{sec-heteroscedastic-noise}. The experimental setup is presented in \autoref{sec-experimental-setup}, followed by results and analysis in \autoref{sec-results-and-analysis}. Finally, we provide concluding remarks in \autoref{sec-conclusion}. 

\section{Related Work}
\label{sec-related-work}
In recent related work, we have proposed sparse Gaussian Processes for photometric redshift inference \citep{almosallam2015}. Gaussian Processes (GPs) are very powerful probabilistic models for regression that are easy to implement. However, the quadratic storage cost and the cubic computational complexity required to train them is deemed impractical for many applications where scalability is a major concern; requiring far more efficient approximations. One of the most common approximations used for GPs is to reduce the computational cost required to invert the $n\times n$ covariance matrix (which gives GPs their computational complexity), where $n$ is the number of samples in the training set. One can also take advantage of the structure of the covariance matrix, if the recordings are evenly spaced in a time series problem for example, then the covariance matrix will have a Toeplitz structure which can be inverted much faster \citep{zhang2005time}. Another approach is to decompose the covariance matrix as a sum of Kronecker products to simplify the computation of the inverse \citep{tsiligkaridis2013}. These properties do not always hold; however, the covariance matrix will always be a positive semi-definite matrix which one can exploit to compute a good approximation of the inverse by treating the problem as a system of linear equations and use the conjugate gradient (CG) method to solve it \citep{gibbs97}. However, the inverse needs to be computed several times during the optimization process of the internal parameters and providing an approximate inverse to the optimizer will cause it to be unstable.

A second class of approaches is to reduce the size of the covariance matrix by means of sparse approximations, instead of using the entire $n$ samples in the training set, a set of $m \ll n$ samples are used to construct the covariance matrix. The samples can be pre-selected either randomly or in an unsupervised fashion such as in \citet{foster2009} where the active set is selected to increase the stability of the computation. Alternatively, one may search for ``pseudo'' points not necessarily present in the training set (and not necessarily even lying within the data range) to use as the active set such that the probability of the data being generated from the model is maximized \citep{snelson2005}. This approach uses a richer likelihood that models input-dependent heteroscedastic noise. However, it assumes a specific form for the noise process and uses a global kernel definition. A comprehensive overview of sparse approximation methods is detailed in \citet{candela2005}. We provide a full formal description of GPs in \autoref{append-gaussian-processes} of the appendix and the reader is advised to read \citet{rasmussen2006gaussian} for a complete review of GPs. Except for \citet{snelson2005}, none of the previously discussed methods account for variable noise, with variations in the posterior variance estimates providing an indication of the model's confidence about its mean function, not the noise, due to the underlying assumption that the observed data has constant white Gaussian noise. One method to learn heteroscedastic noise is to model both the mean and the noise functions as Gaussian Processes. This is achieved by first holding the noise fixed and optimizing with respect to the mean, then holding the mean fixed and optimizing with respect to the noise and repeated until convergence \citep{kersting2007most}, this can be viewed as a group-coordinate ascent optimization. In this paper, we use basis function models (BFM), viewed as a sparse GP method, and provide novel methods to enhance the posterior variance accuracy. 

\section{Sparse Gaussian Processes}
\label{sec-sparse-gaussian-processes}
In this section, we describe sparse Gaussian processes as basis function models, whose semi-parametric form is defined via a set of weights. The underlying assumption in a BFM is that, given a set of inputs $\mathbfss{X}=\left\{\mathbfit{x}_{i}\right\}_{i=1}^{n}\in \mathbb{R}^{n\times d}$ and a set of target outputs $\mathbfit{y}=\left\{y_{i}\right\}_{i=1}^{n}\in \mathbb{R}^{n}$, where $n$ is the number of samples in the data set and $d$ is the dimensionality of the input, that the observed target $y_{i}$ is generated by a linear combination of $m$ non-linear functions $\bphi\left(\mathbfit{x}_{i}\right)=\left[\phi_{1}\left(\mathbfit{x}_{i}\right),\hdots,\phi_{m}\left(\mathbfit{x}_{i}\right)\right]\in \mathbb{R}^{m}$ of the input plus additive noise $\epsilon_{i}\sim\mathcal{N}\left(0,\beta^{-1}\right)$:

\begin{align}
y_{i} = \bphi\left(\mathbfit{x}_{i}\right)\mathbfit{w}+\epsilon_{i},
\end{align}
where $\mathbfit{w}$ is a vector of length $m$ of real-valued coefficients, or the parameters of the model. In the case of photometric redshift estimation, $\mathbfss{X}$ are the photometric measurements and associated uncertainties of the filters, namely $d$ inputs, and $n$ training objects.

Throughout the rest of the paper, $\mathbfss{X}\left[i,:\right]$ denotes the $i$-th row of matrix $\mathbfss{X}$, or $\mathbfit{x}_{i}$ for short, whereas $\mathbfss{X}\left[:,j\right]$ denotes the $j$-th column, $\mathbfss{X}\left[i,j\right]$ denotes the element at the $i$-th row and the $j$-th column in matrix $\mathbfss{X}$, and similarly for other matrices. Note that the mean of the predictive distribution derived from a GP is a linear combination of $n$ kernel functions. The BFM approach is to assume the form of the function to be a linear combination of $m\ll n$ \emph{basis} functions and integrates out its parameters. In this paper, we choose the radial basis function (RBF) kernel as our basis function, defined as follows:

\begin{align}
\phi_{j}\left(\mathbfit{x}_{i}\right)=\exp\left(-\frac{1}{2}\left(\mathbfit{x}_{i}-\mathbfit{p}_{j}\right)^{T}\Gamma_{j}^{T}\Gamma_{j}\left(\mathbfit{x}_{i}-\mathbfit{p}_{j}\right)\right),
\label{eq-gpgl-rbf}
\end{align}
where we define $\mathbfss{P}=\left\{\mathbfit{p}_{i}\right\}_{i=1}^{m}\in \mathbb{R}^{m\times d}$ to be the set of basis vectors associated with the basis functions and $\Gamma_{j}^{T}\Gamma_{j}$, $\Gamma_{j}\in\mathbb{R}^{d\times d}$, are bespoke precision matrices associated with each basis function. We refer to the model with such basis functions as Gaussian processes with variable covariances, or {\sc gpvc}. The framework also allows for other types of covariance structures, the options include:

\begin{itemize}  
\item {\sc gpvc}: Variable covariances, or a bespoke $\Gamma_{j}$ for each basis function $j$.
\item {\sc gpgc}: A global covariance, or a shared $\Gamma$ for all basis functions.
\item {\sc gpvd}: Variable diagonal covariances, or a bespoke \emph{diagonal} $\Gamma_{j}$ for each basis function $j$.
\item {\sc gpgd}: A global diagonal covariance, or a shared \emph{diagonal} $\Gamma$ for all basis functions.
\item {\sc gpvl}: Variable length-scales, or a bespoke \emph{isotropic} covariance for each basis function $j$; i.e. $\Gamma_{j}=\mathbfss{I}\gamma_{j}$, where $\gamma_{j}$ is a scaler.
\item {\sc gpgl}: A global length-scale, or a shared \emph{isotropic} covariance for all basis functions$\Gamma=\mathbfss{I}\gamma$, where $\gamma$ is a scaler.
\end{itemize}

We assume, for now, that our observations are noisy with a constant precision $\beta$ and a mean of zero. This is obviously not true in reality as the photometric noise is dependent on the depth of the individual images in each band. Note that these assumptions are made to simplify our illustration and we relax these later in the paper. The likelihood is hence defined as follows:

\begin{align}
p\left(\mathbfit{y}|\mathbfit{w}\right) =& \mathcal{N}\left(\Phi\mathbfit{w},\beta^{-1}\mathbfss{I}\right),\\
\Phi=&\begin{bmatrix}\bphi\left(\mathbfit{x}_{1}\right)\\\vdots\\\bphi\left(\mathbfit{x}_{n}\right)\end{bmatrix}.
\end{align}
We now need to define a prior on $\mathbfit{w}$ in order to proceed. We use a prior that promotes a smooth function, hence preferring the simplest explanation that fits the data. The smoothness assumption also transforms the objective from an ill-posed problem to a well-posed one, as there are an infinite number of functions that would fit the data. This can be achieved by requiring the weights in $\mathbfit{w}$ to be independent and the norm as small as possible. This can be formulated probabilistically by taking $p\left(\mathbfit{w}\right)=\mathcal{N}\left(0,\alpha^{-1}\right)$, where $\alpha$ is the prior precision of the parameters $\mathbfit{w}$. With a likelihood and a prior, we can derive the posterior as $p\left(\mathbfit{w}|\mathbfit{y}\right)=p\left(\mathbfit{y}|\mathbfit{w}\right)p\left(\mathbfit{w}\right)/p\left(\mathbfit{y}\right)$ from Bayes theorem, which can be shown to have the following normal distribution \citep{bishop2006}:

\begin{align}
p\left(\mathbfit{w}|\mathbfit{y}\right) =& \mathcal{N}\left(\bar{\mathbfit{w}},\Sigma\right),\\
\bar{\mathbfit{w}} =&\beta\Sigma^{-1}\Phi^{T}\mathbfit{y},\\
\Sigma =& \beta\Phi^{T}\Phi+\alpha\mathbfss{I}.
\end{align}

The marginal likelihood, or the evidence function \citep{bishop2006}, can be derived by integrating out $\mathbfit{w}$ as in \autoref{eq-integrate-w}.
\begin{align}
p\left(\mathbfit{y}\right)=\int p\left(\mathbfit{y}|\mathbfit{w}\right)p\left(\mathbfit{w}\right)\mbox{d}\mathbfit{w}.
\label{eq-integrate-w}
\end{align}
This can be expressed in terms of the mean $\bar{\mathbfit{w}}$ and the covariance $\Sigma$ of the posterior distribution:
\begin{align}
\ln p\left(\mathbfit{y}\right) =&-\frac{\beta}{2} \left\|\Phi\bar{\mathbfit{w}}-\mathbfit{y}\right\|^{2}+\frac{n}{2}\ln\beta-\frac{n}{2}\ln\left(2\pi\right)\label{eq-log-marginal-bfm}\\
&-\frac{\alpha}{2} \bar{\mathbfit{w}}^{T}\bar{\mathbfit{w}}+\frac{m}{2}\ln\alpha-\frac{1}{2}\ln|\Sigma|.\nonumber
\end{align}
The hyperparameters of the basis function, the precision $\beta$, the weight precision $\alpha$ and the pseudo points' locations $\mathbfss{P}$ can now be optimized with respect to the log marginal likelihood defined in \autoref{eq-log-marginal-bfm}. Once the parameters have been inferred, the predictive distribution of an unseen test case $\mathbfit{x}_{*}$ is distributed as follows \citep{bishop2006}:

\begin{align}
p\left(y_{*}|\mathbfit{y}\right)=& \mathcal{N}\left(\mu_{*},\sigma_{*}^{2}\right),\\
\mu_{*} =& \bphi\left(\mathbfit{x}_{*}\right)\bar{\mathbfit{w}},\\
\sigma_{*}^{2} =&\nu_{*}+\beta^{-1},\label{eq-sigma}\\
\nu_{*} =&\bphi\left(\mathbfit{x}_{*}\right)\Sigma^{-1}\bphi\left(\mathbfit{x}_{*}\right)^{T},\label{eq-nu}
\end{align}
Note that we are no longer restricted to Mercer kernels or a single basis function definition. The basis function can therefore be modelled using variable length-scales and variable covariances as in \citet{almosallam2015}, to capture different kinds of patterns that can arise in different regions of the input space. Basis function models can be shown to be a degenerate GP with an equivalent kernel function $\kappa\left(\mathbfit{x}_{i},\mathbfit{x}_{j}\right)=\alpha\bphi\left(\mathbfit{x}_{i}\right)\bphi\left(\mathbfit{x}_{j}\right)^{T}$ \citep{candela2005}.

\section{Automatic Relevance Determination}
\label{sec-automatic-relevance-determination}
In addition to achieving accurate predictions we wish to minimize the number of basis functions, to produce a sparse model representation. Instead of adding an additional prior over the number of basis functions, we can achieve this goal by incorporating a sparsity-inducing prior on $\mathbfit{w}$. We use prior diagonal precision matrix $\mathbfss{A}=\mbox{diag}(\balpha)$, where $\balpha=\left\{\alpha_{i}\right\}_{i=1}^{m}$, or a precision parameter per weight. The modified prior is $p\left(\mathbfit{w}\right)=\mathcal{N}\left(0,\mathbfss{A}^{-1}\right)$ and the log marginal likelihood is simply extended as follows:

\begin{align}
\ln p\left(\mathbfit{y}\right) =&-\frac{\beta}{2} \left\|\Phi\bar{\mathbfit{w}}-\mathbfit{y}\right\|^{2}+\frac{n}{2}\ln\beta-\frac{n}{2}\ln 2\pi\\
&-\frac{1}{2} \bar{\mathbfit{w}}^{T}\mathbfss{A}\bar{\mathbfit{w}}+\frac{1}{2}\ln\left|\mathbfss{A}\right|-\frac{1}{2}\ln|\Sigma|,\label{eq-log-marginal-bfm-ard}\nonumber\\
\mbox{where }\Sigma=&\beta\Phi^{T}\Phi+\mathbfss{A}.
\end{align}
By modelling each weight with its associated precision, we enable a natural shrinkage (or regularization). Take, for example, a specific precision $\alpha_{i}$; note that maximizing $\frac{1}{2}\ln\alpha_{i}$ will minimize $-\frac{1}{2}\bar w_{i}^{2}\alpha_{i}$, unless $\bar w_{i}=0$. The optimization routine will therefore drive as many of the weights to zero as possible thus maintaining the least number of basis functions relevant to model the data. A similar approach was proposed by \citet{tipping2001} coined as the relevance vector machine (RVM), where the set of basis function locations $\mathbfss{P}$ was set equal to the locations of the training samples $\mathbfss{X}$ and held fixed. Only the precision parameter $\beta$ and the $\balpha$ values were optimized to determine the relevant set of vectors from the training set. This approach is still computationally expensive and \cite{tipping2001} proposed an iterative workaround to add and remove vectors incrementally. 

\section{Heteroscedastic Noise}
\label{sec-heteroscedastic-noise}
The predictive variance in \autoref{eq-sigma} has two components, the first term $\nu_{*}$ is the model variance and the second term $\beta^{-1}$ is the noise uncertainty. The model variance thus depends on the data density of the training sample at $\mathbfit{x}_{*}$. Theoretically, this component of the model variance will go to zero as the size of the data set increases. This term hence models our underlying uncertainty about the mean function. The model becomes very confident about the posterior mean when presented with a large number of samples at $\mathbfit{x}_{*}$, or in photometric redshift terms in a particular region of colour-redshift space, at which point the predictive variance reduces to the intrinsic noise variance. The latter, at this point, is assumed to be white Gaussian noise with a fixed precision $\beta$.

In this section, we enhance the model's predictive variance estimation by modelling the noise variance as a function of input, or $\beta_{i}=f\left(\mathbfit{x}_{i}\right)$ to account for variable and input-dependent noise, i.e. heteroscedastic noise, as is the case for imaging using different surveys. We choose to model the function as a linear combination of basis functions via $\beta_{i}=\exp\left(\bphi\left(\mathbfit{x}_{i}\right)\mathbfit{u}+b\right)$, where we choose the exponential form to ensure positivity of $\beta_{i}$. Note that if $\mathbfit{u}=\mathbfit{0}$ and $b=\ln\beta$, the model reduces to the original assumption of a fixed precision $\beta$. We thus redefine the likelihood as follows:
\begin{align}
p\left(\mathbfit{y}|\mathbfit{w}\right) = \mathcal{N}\left(\Phi\mathbfit{w},\mathbfss{B}^{-1}\right),
\end{align}
where $\mathbfss{B}$ is a $n\times n$ diagonal matrix where each element across the diagonal $\mathbfss{B}\left[i,i\right]=\beta_{i}$. Following the same procedure, the posterior $p\left(\mathbfit{w}|\mathbfit{y}\right)$ is expressed as follows:
\begin{align}
p\left(\mathbfit{w}|\mathbfit{y}\right) =& \mathcal{N}\left(\bar{\mathbfit{w}},\Sigma\right),\\
\bar{\mathbfit{w}} =& \Sigma^{-1}\Phi^{T}\mathbfss{B}\mathbfit{y},\\
\Sigma =& \Phi^{T}\mathbfss{B}\Phi+\mathbfss{A},
\end{align}
and the updated log marginal likelihood becomes:
\begin{align}
\ln p\left(\mathbfit{y}\right) =&-\frac{1}{2} \bdelta^{T}\mathbfss{B}\bdelta+\frac{1}{2}\ln\left|\mathbfss{B}\right|-\frac{n}{2}\ln 2\pi\\
&-\frac{1}{2} \bar{\mathbfit{w}}^{T}\mathbfss{A}\bar{\mathbfit{w}}+\frac{1}{2}\ln\left|\mathbfss{A}\right|-\frac{1}{2}\ln|\Sigma|,\label{eq-log-marginal-var}\nonumber
\end{align}
where $\bdelta=\Phi\bar{\mathbfit{w}}-\mathbfit{y}$. Note that cost-sensitive learning \citep{almosallam2015}, can be readily incorporated into our model by setting $\mathbfss{B}\left[i,i\right]=\beta_{i}\omega_{i}$, where $\omega_{i}=\left(1+ z_{i}\right)^{-2}$, in which $z_{i}$ is the spectroscopic redshift for source $i$. In addition, we also add a prior on $\mathbfit{u}$ to favour the simplest precision function, namely that $\mathbfit{u}$ is normally distributed with a mean of 0 and a diagonal precision matrix $\mathbfss{N}=\mbox{diag}(\boldeta)$, or $\mathbfit{u}\sim\mathcal{N}\left(0,\mathbfss{N}^{-1}\right)$, where $\boldeta=\left\{\eta_{i}\right\}_{i=1}^{m}$. The final objective function to be optimized is thus the log marginal likelihood plus the log of the prior on $\mathbfit{u}$,
\begin{align}
\ln p\left(\mathbfit{y}\right) =&-\frac{1}{2} \bdelta^{T}\mathbfss{B}\bdelta+\frac{1}{2}\ln\left|\mathbfss{B}\right|-\frac{n}{2}\ln 2\pi\\
&-\frac{1}{2} \bar{\mathbfit{w}}^{T}\mathbfss{A}\bar{\mathbfit{w}}+\frac{1}{2}\ln\left|\mathbfss{A}\right|-\frac{1}{2}\ln|\Sigma|\nonumber\\
&-\frac{1}{2} \mathbfit{u}^{T}\mathbfss{N}\mathbfit{u}+\frac{1}{2}\ln\left|\mathbfss{N}\right|-\frac{m}{2}\ln 2\pi.\nonumber
\end{align}
The parameter $\boldeta$ hence acts as an automatic relevance determination cost for the noise process, allowing the objective to dynamically select different sets of relevant basis functions for both the posterior mean and variance estimation. The probability of unseen test cases is normally distributed as follows:
\begin{align}
p\left(y_{*}|\mathbfit{y}\right)=& \mathcal{N}\left(\mu_{*},\sigma_{*}^{2}\right),\\
\mu_{*} =& \bphi\left(\mathbfit{x}_{*}\right)\bar{\mathbfit{w}},\\
\sigma_{*}^{2} =&\nu_{*}+\beta_{*}^{-1},\label{eq-variable-sigma}\\
\beta_{*}=&\exp\left(\bphi\left(\mathbfit{x}_{*}\right)\mathbfit{u}+b\right),\label{eq-beta}
\end{align}
where $\beta_{*}^{-1}$ is the input-dependent noise uncertainty and $\nu_{*}$ is defined in \autoref{eq-nu}. It is worth mentioning that in the parameter space, $\mathbfit{w}$, the problem is convex; and thus can be modelled using a single Gaussian distribution. In the hyper-parameter space however, the problem can be highly non-convex with many local minima. This adds an extra source of uncertainty about the model due to training that is dependent on the initial condition and the optimization procedure. This can be addressed using a committee of models, where each model is initialized differently, to fit a mixture of Gaussian distribution instead of a single one to better fit the true model distribution \citep{penny1997,roberts1996}.

We search for the optimal set of model parameters using a gradient-based optimization; hence we require the derivatives of the log marginal likelihood with respect to each parameter. The gradient calculations are provided in section \autoref{append-optimization} of the appendix, for both the general case of any basis function and an efficient procedure for the six different configurations of RBFs. In this paper, the Limited-memory Broyden-Fletcher-Goldfarb-Shanno algorithm (LBFGS) is used to optimize the objective. This uses a quasi-Newton method to compute the search direction in each step by approximating the inverse of the Hessian matrix from the history of gradients in previous steps \citep{jorge1980}. We use the {\sc minFunc} optimization toolbox by \cite{schmidt2005}.

\section{Experimental Setup}
\label{sec-experimental-setup}
\subsection{Tested models}
The focus of the method described in this paper is to generate input-dependent predictive distributions, we therefore only include photometric redshift algorithms from the literature that produce point estimates of the posterior mean (the expected value), as well as uncertainty predictions (typically a predictive variance) for each source, given its photometry. For comparison, we test our proposed approach against {\sc annz2} \citep{Sadeh2015}, {\sc tpz} \citep{kind2013} and Sparse Pseudo-Input Gaussian Processes ({\sc spgp}) which also generate uncertainty predictions. {\sc annz2} is an extension of {\sc annz}, a popular artificial neural network (ANN) based code \citep{Collister04}. {\sc annz2} utilizes many machine learning methods (MLM) including ANNs, decision trees and $k$-nearest neighbours (kNN). {\sc annz2} can be considered as a committee machine that combines the results of different models with various configurations, initializations and optimization techniques. For instance, the output of many ANNs with different number of layers, number of hidden units, input preprocessing, number of trees and sampling methods. {\sc tpz} is a random forest implementation that generates predictions by subdividing the data based on its features until a termination leaf is reached, determined using an information gain metric that measures the information quality of each feature and its ability to predict the desired output. The algorithm generates a number of trees, each trained on a subsample of features, which proves to be more effective and stable than a single tree trained on all features. {\sc spgp} is a sparse GP model that uses pseudo-inputs as the basis set to determine the covariance function of the GP \citep{snelson2005}. The pseudo-inputs are treated as parameters of the model that are optimized to maximize the log marginal likelihood. {\sc spgp} is similar to the {\sc gpgl} model except that the prior covariance of $\mathbfit{w}$, is set to the covariance matrix of the pseudo-inputs, instead of setting it to $\mathbfss{A}^{-1}$. The posterior variance is inferred using a stationary noise model in {\sc spgp}, whereas the posterior variance in {\sc gpgl} is modelled as a separate function of the basis, hence allowing for non-stationarity and input sensitivity.

\subsection{The data set}
We train the models on the Sloan Digital Sky Survey's 12th Data Release \citep[SDSS;][]{SDSS3}. We select galaxies where both the photometry and the spectroscopic redshifts are available. The total number of sources is 2,120,465, which contains 1,301,943 from the Baryon Oscillation Spectroscopic Survey (BOSS), 817,657 from the SDSS-III survey, 826 from Segue-1 and 93 from Segue-2. The {\fontfamily{pcr}\selectfont modelMag} magnitudes for the $u$,$g$,$r$,$i$ and $z$ bands were used with their associated error estimates. We preprocess the associated uncertainties of the photometry by replacing them with their natural log to transform the domain of the features from the positive domain to the real domain. This has the advantage of having all the features share the same domain and allows for a fully unconstrained optimization. In addition, we use principle component analysis \citep[PCA;][]{jolliffe1986} to de-correlate the features such that the data have a zero mean and an identity covariance, but retain all features with no dimensionality reduction. Decorrelation speeds up the optimization processes and offers some numerical advantages. This approach, often referred to as `sphering' or `whitening' in the literature, is a common practice \citep{bishop2006}. We randomly sampled three sets of 100,000 sources each for training, validation and testing. The training set was used for learning the model, the validation set for model selection and the test set to report the results. The SQL statement used to create the data set is provided in \autoref{append-sql-statement} of the appendix.

\subsection{Metrics}
Four metrics are considered to compare the results of the different methods. The mean log likelihood (MLL), the root mean squared error (RMSE), the fraction retained (FR) which provides a metric for the level of catastrophic outliers from the one-to-one relation, and the bias. These are defined as below:

\begin{align}
\text{MLL}=& \frac{1}{n}\sum_{i=1}^{n}-\frac{1}{2\sigma_{i}^{2}}\left(z_{i}-\acute{z}_{i}\right)^{2}-\frac{1}{2}\ln\sigma_{i}^{2}-\frac{1}{2}\ln2\pi,\\
\text{RMSE} =&\sqrt{\frac{1}{n}\sum_{i=1}^{n}\left(\frac{z_{i}-\acute{z}_{i}}{1+z_{i}}\right)^{2}},\\
\text{FR}_{e} =&\frac{100}{n}\left|\left\{i:\left|\frac{z_{i}-\acute{z}_{i}}{1+z_{i}}\right|<e\right\}\right|,\\
\text{Bias} =&\frac{1}{n}\sum_{i=1}^{n}\frac{z_{i}-\acute{z}_{i}}{1+z_{i}}, \label{eq-bias}
\end{align}
where $z_{i}$ is the spectroscopic redshift for source $i$, $\acute{z}_{i}$ is the predicted photometric redshift, $\sigma_{i}^{2}$ is the predicted variance and $e$ is the outlier threshold, i.e. FR$_{0.15}$ is the fraction of samples where $\left|\left(z_{i}-\acute{z}_{i}\right)/\left(1+z_{i}\right)\right|$ is less than 0.15. The log likelihood is a natural way to evaluate the point estimate and the uncertainty prediction at the same time. The first term is the weighted sum of squared errors, where the weights are the predicted variance, prefers larger variance, whereas the second term punishes for large variances on a log scale. The advantage of this form, is that the optimal variance, if everything else were to set fixed, is exactly the squared error. 
\section{Results and Analysis}
\label{sec-results-and-analysis}

In the following, we analyse the results from the various machine learning methods within a number of tests. For the predictive mean, we use the {\fontfamily{pcr}\selectfont zmean1} prediction from {\sc tpz} and the {\fontfamily{pcr}\selectfont ANNZ\_best} score from {\sc annz2}. We use the square of {\fontfamily{pcr}\selectfont err1} from {\sc tpz} and the square of {\fontfamily{pcr}\selectfont ANNZ\_best\_err} from {\sc annz2} as the predictive variance.

\subsection{Model complexity}
\begin{figure*}
        \centering
        \begin{subfigure}[b]{0.45\textwidth}
                 \includegraphics[width=\textwidth]{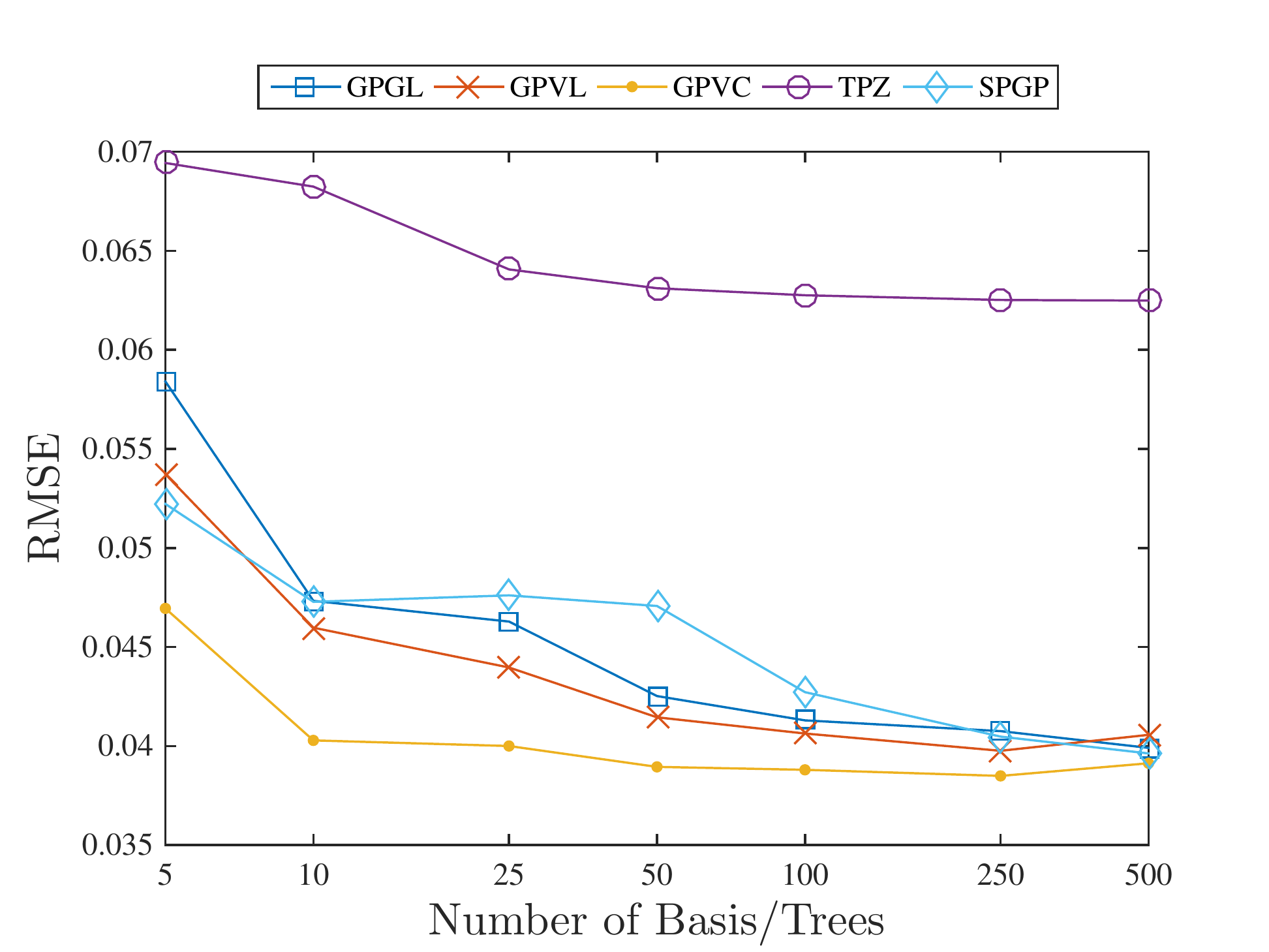}
                 \caption{RMSE}
        \end{subfigure}
        ~
        \begin{subfigure}[b]{0.45\textwidth}
                 \includegraphics[width=\textwidth]{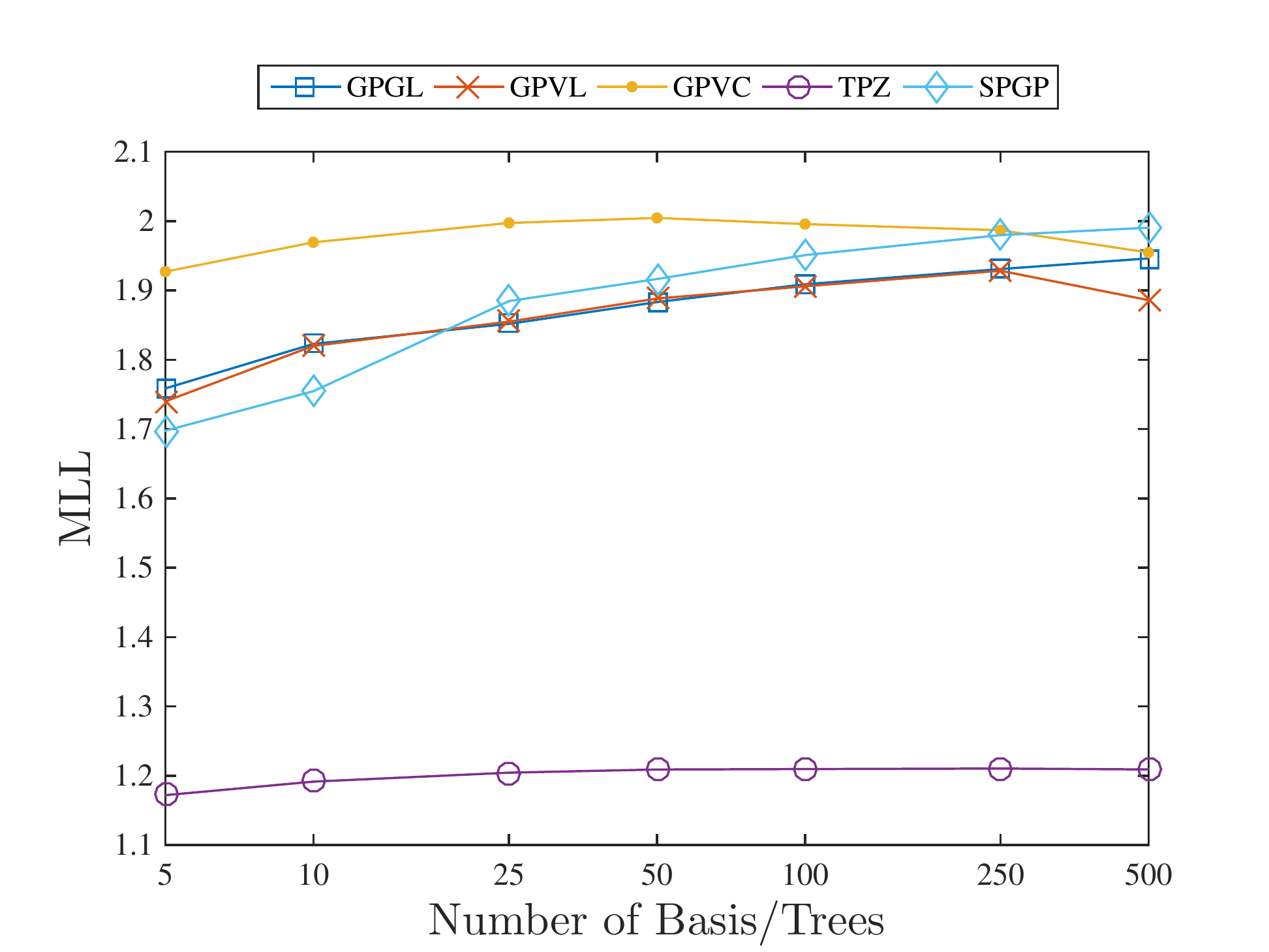}
                 \caption{MLL}
        \end{subfigure}
        
        \begin{subfigure}[b]{0.45\textwidth}
                 \includegraphics[width=\textwidth]{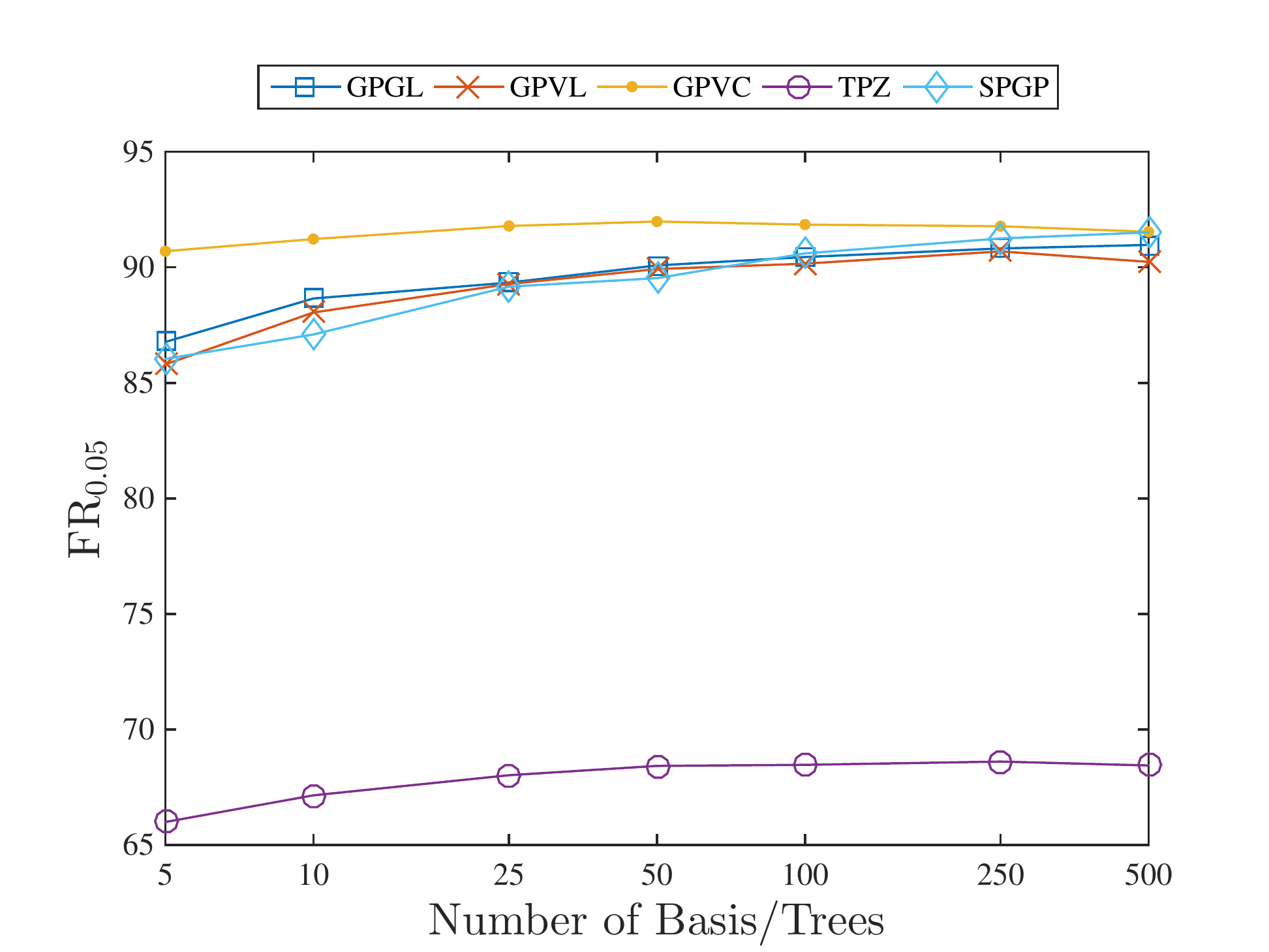}
                 \caption{FR$_{0.05}$}
        \end{subfigure}
        ~
        \begin{subfigure}[b]{0.45\textwidth}
                 \includegraphics[width=\textwidth]{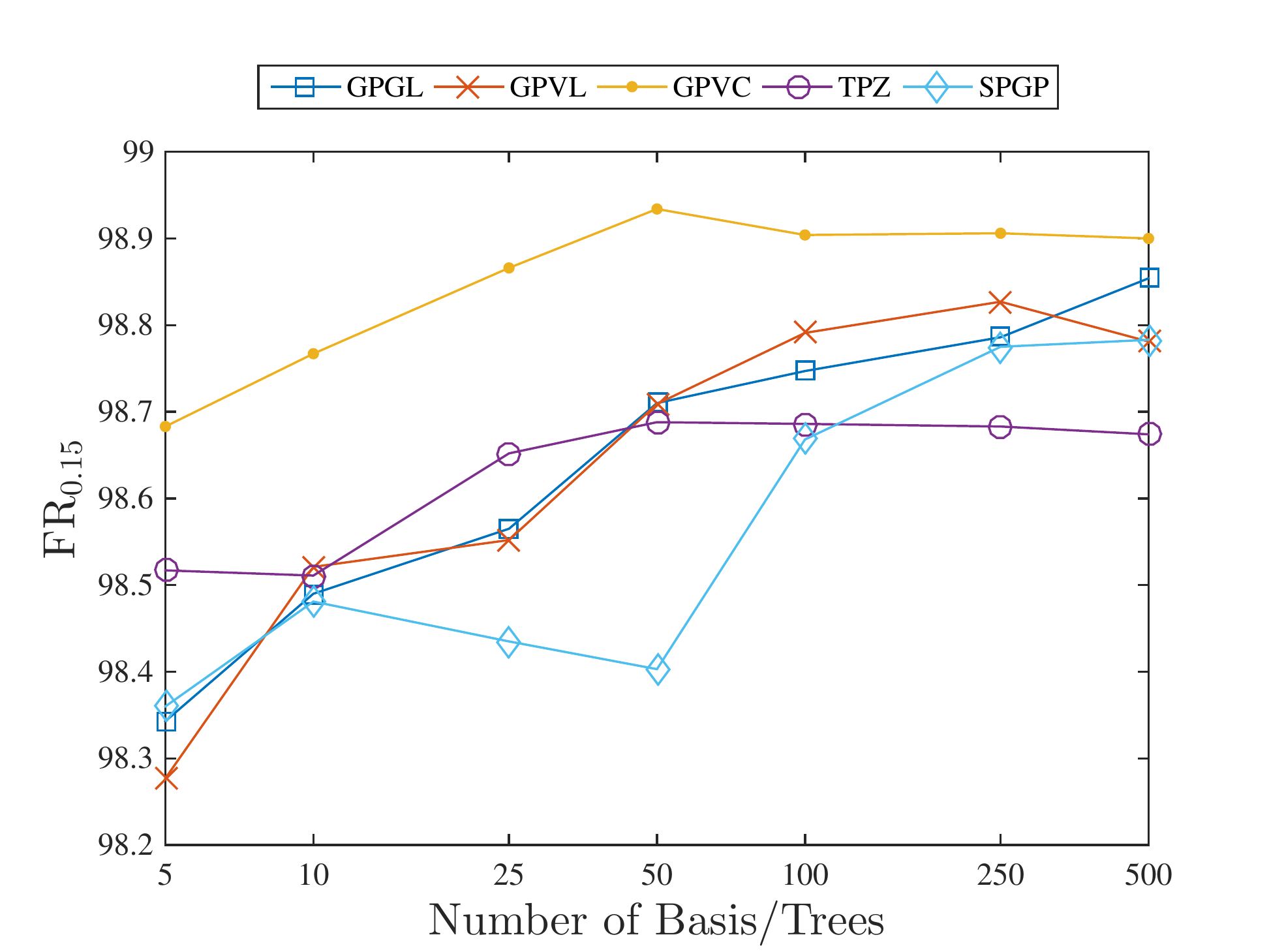}
                 \caption{FR$_{0.15}$}
        \end{subfigure}
        
       \caption{The (a) RMSE, (b) MLL, (c) FR$_{0.05}$ and (c) FR$_{0.15}$ performance of each method on the test set using different numbers of basis/trees.}
	\label{fig-model-complexity}
\end{figure*}

In the first experiment, we analyse the relationship between the algorithms' complexity and their fit, as measured by the proposed metrics. For {\sc gpgl}, {\sc gpvl}, {\sc gpvc} and {\sc spgp} we vary the number of basis functions, whereas in {\sc tpz} we vary the number of trees in the forest and fix the number of sub-features selected for each tree to the suggested value of $\sqrt{d}\simeq 4$ and keep the remaining configuration options as suggested, since the code is configured for SDSS-like surveys. We tested the models on 5, 10, 25, 50, 50, 100, 250 and 500 basis/trees. {\sc annz2} is an aggregation of many models with various configurations so it is not included in this experiment. \autoref{fig-model-complexity} shows the performance of the methods on the held-out test set as we vary the number of basis/trees. {\sc gpvc} consistently outperforms the other methods in all metrics, reaching a RMSE $\sim 0.039$, FR$_{0.05} \sim 91.9$ per cent and FR$_{0.15} \sim 98.9$ per cent, {\sc tpz} on the other hand is significantly worse in all metrics (RMSE $\sim 0.063$; FR$_{0.05} \sim 68.5$ per cent; FR$_{0.15} \sim 98.7$ per cent).

\subsection{Performance analysis}
\begin{figure}
	\centering
	\includegraphics[width=\columnwidth]{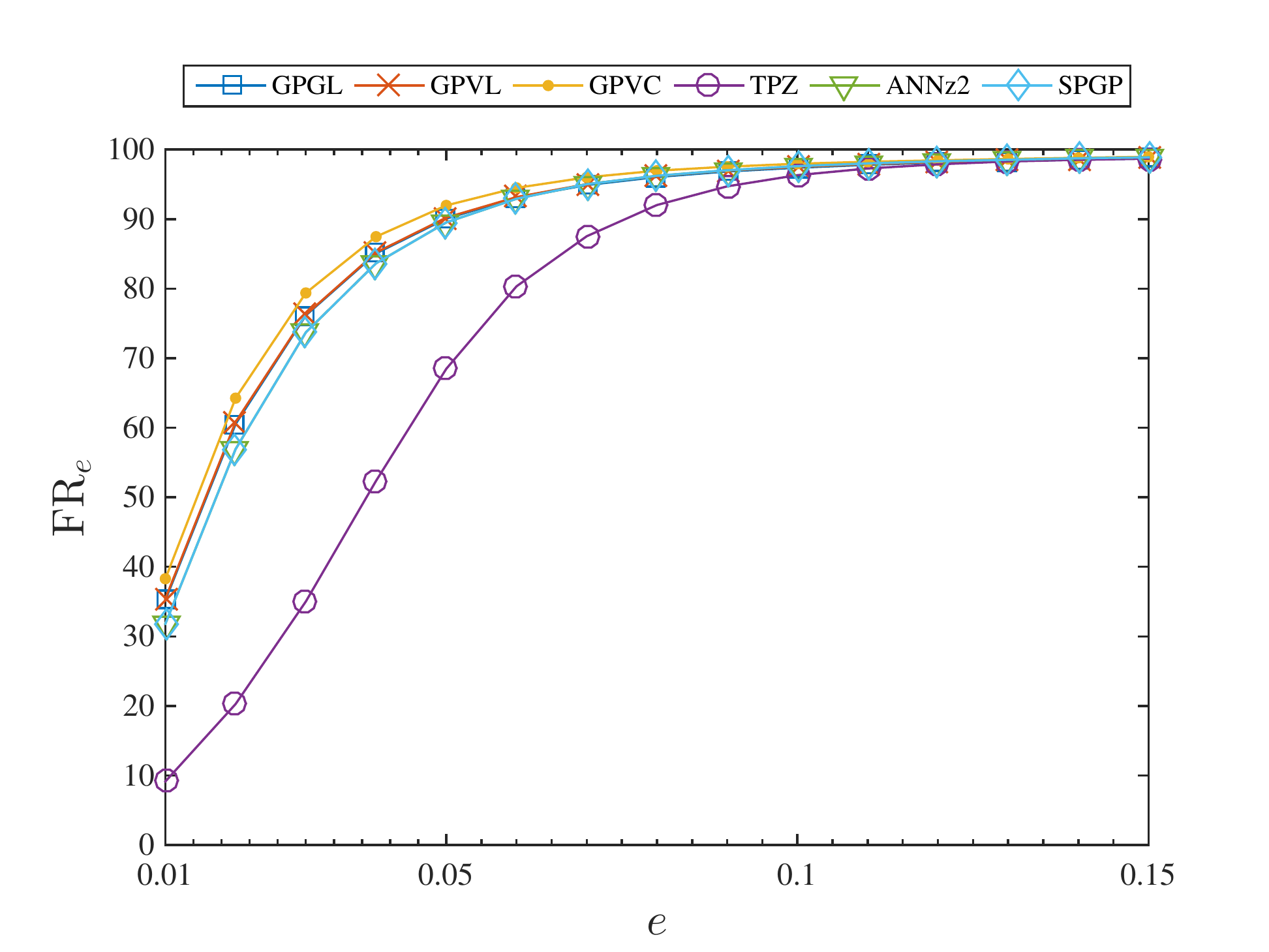}
	\caption{The NAMD$_{e}$ for different values of $e$ for each method using 100 basis/trees/MLMs on the test set.}
	\label{fig-FRe}
\end{figure}

\begin{figure*}
        \centering
        \begin{subfigure}[b]{0.45\textwidth}
                 \includegraphics[width=\textwidth]{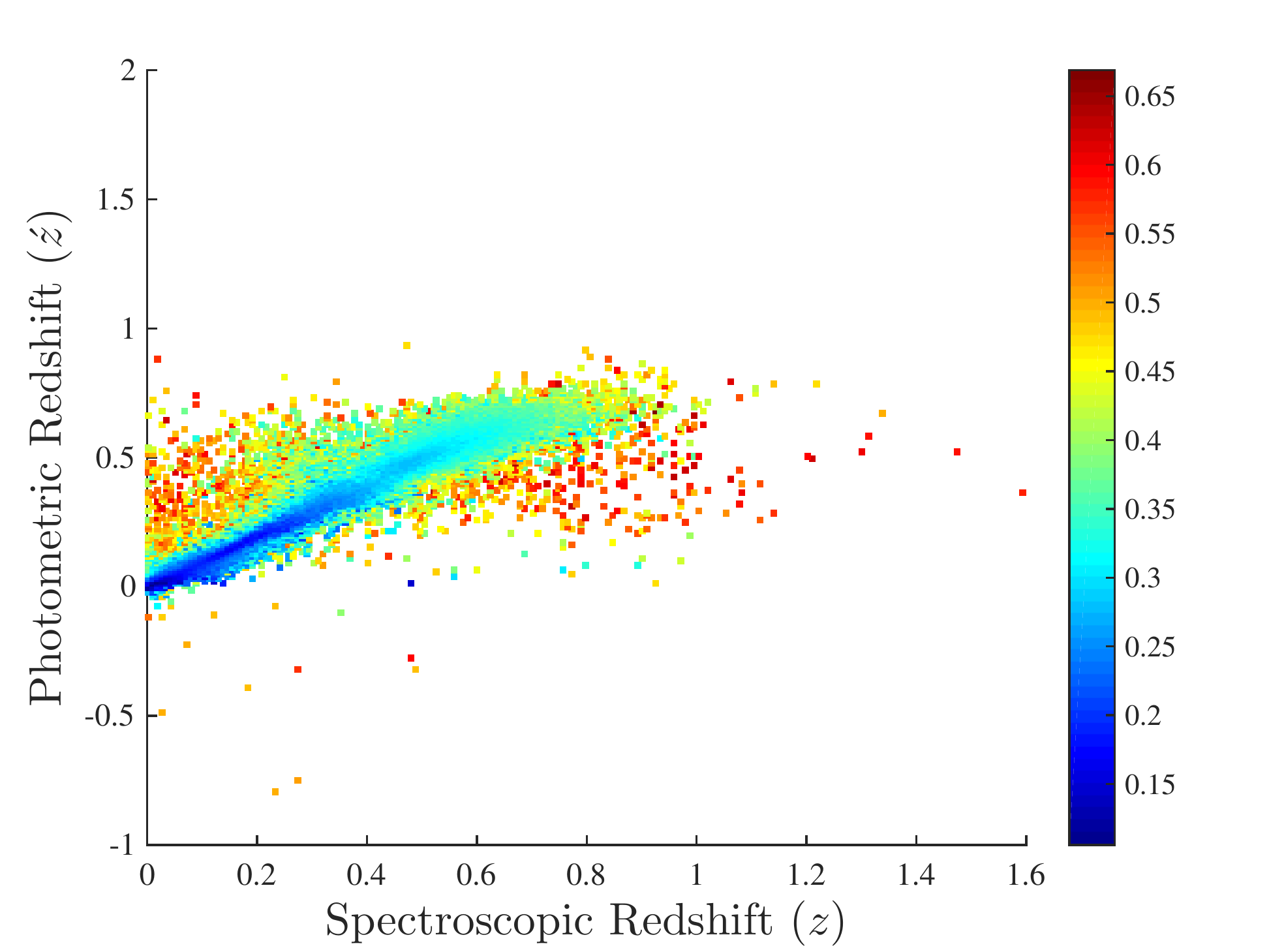}
                 \caption{{\sc gpvc}}
        \end{subfigure}
        ~
        \begin{subfigure}[b]{0.45\textwidth}
                 \includegraphics[width=\textwidth]{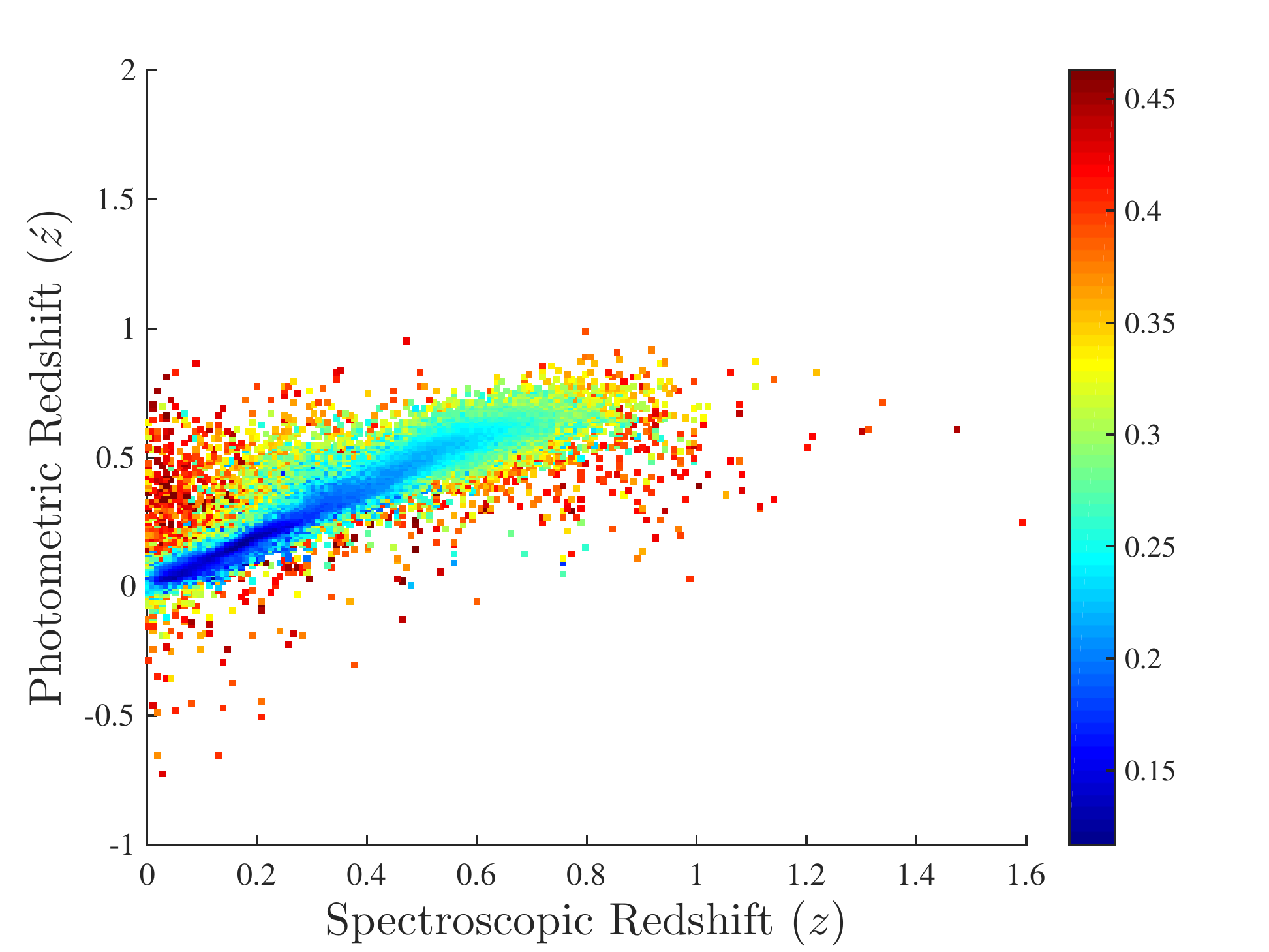}
                 \caption{{\sc spgp}}
        \end{subfigure}
        
        \begin{subfigure}[b]{0.45\textwidth}
                 \includegraphics[width=\textwidth]{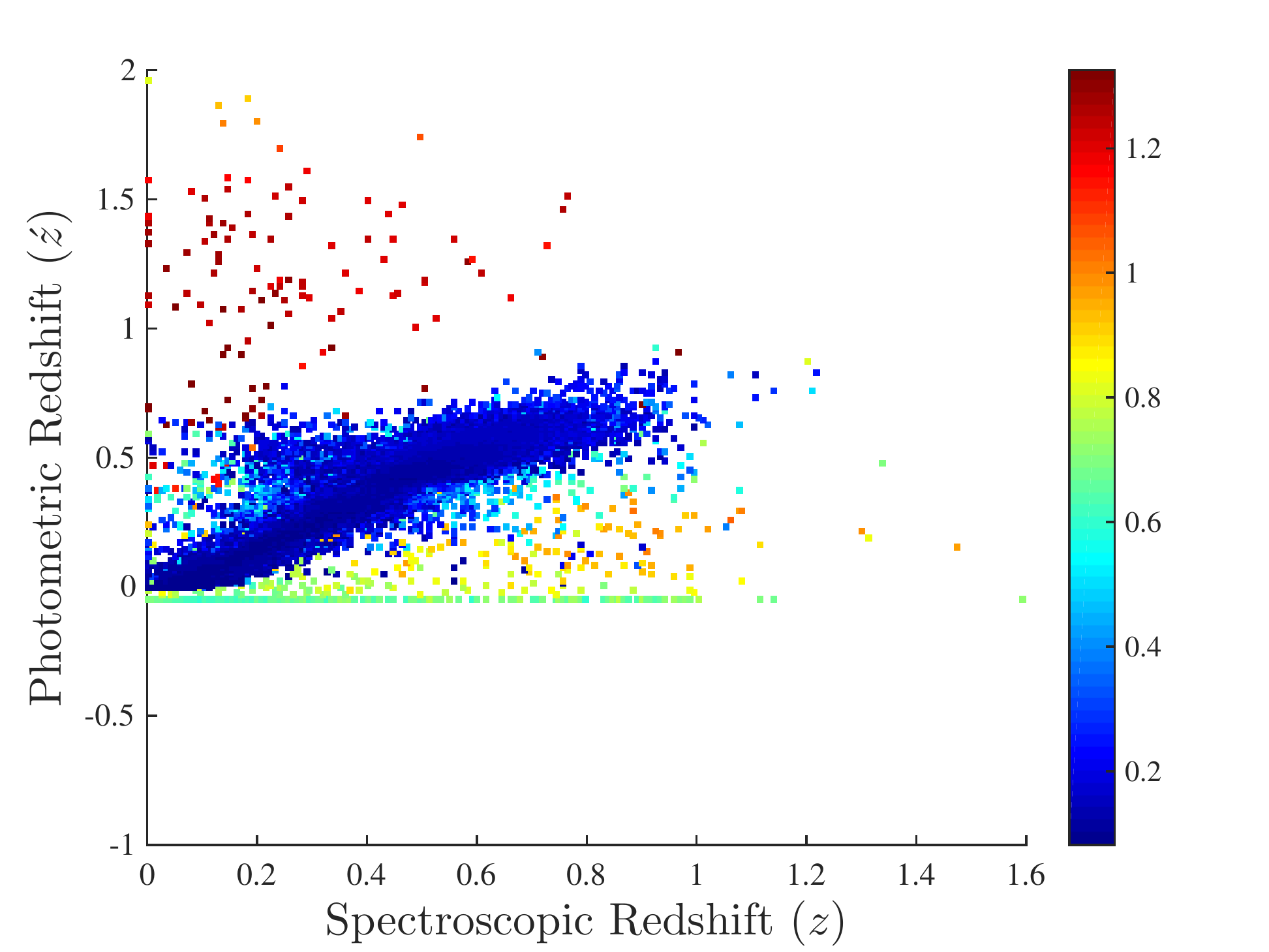}
                 \caption{{\sc tpz}}
        \end{subfigure}
        ~
        \begin{subfigure}[b]{0.45\textwidth}
                 \includegraphics[width=\textwidth]{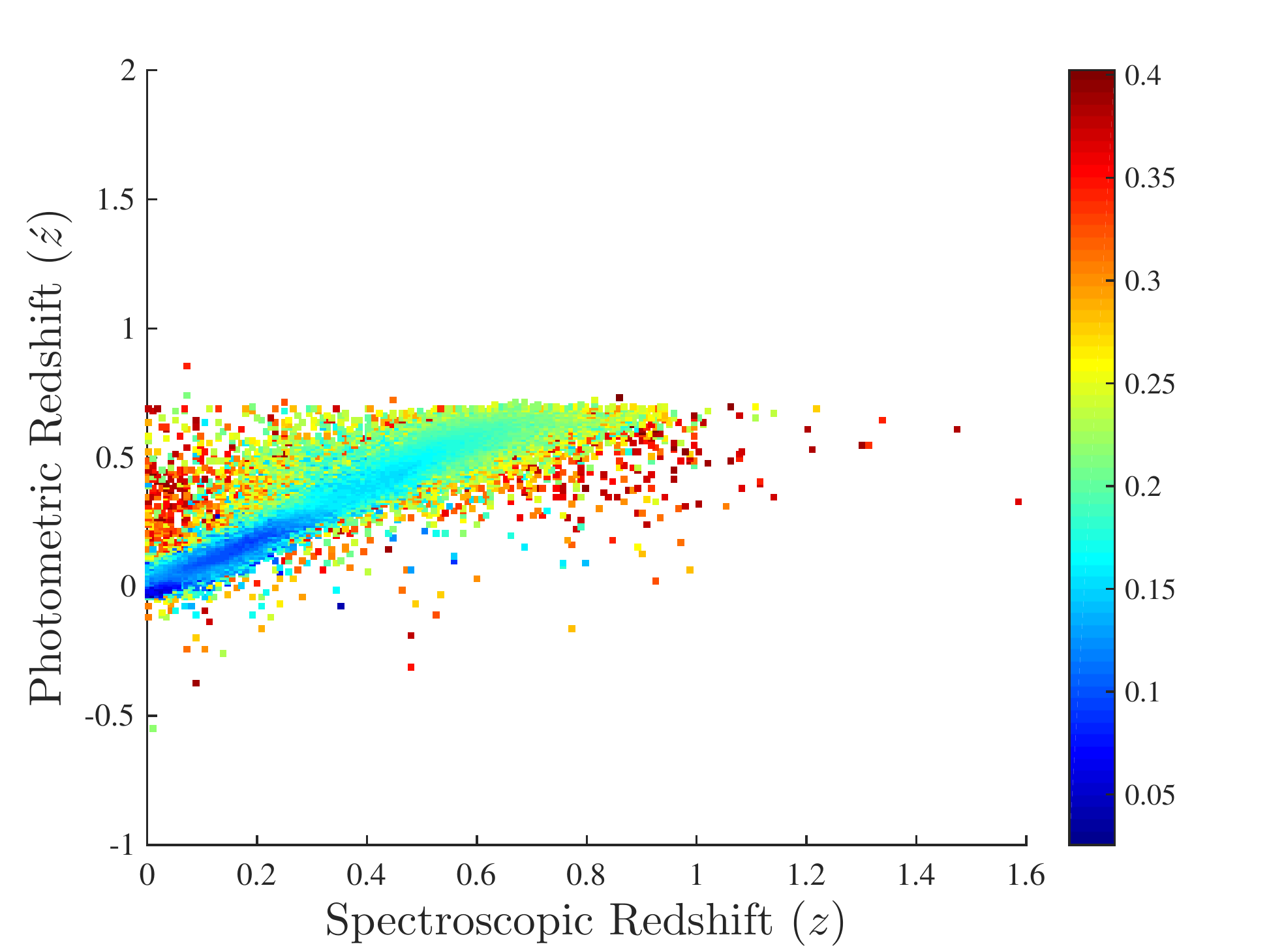}
                 \caption{ {\sc annz2}}
        \end{subfigure}

       \caption{The scatter plots of the spectroscopic redshift $z$ vs. the predicted photometric redshift $\acute{z}$ on the test set for (a) {\sc gpvc}, (b) {\sc spgp}, (c) {\sc tpz} and (d) {\sc annz2} using 100 basis/trees/MLMS. The predictive variance is colour coded, on a log scale, by the value of $\sigma_{*}$ (\autoref{eq-variable-sigma}).}
	\label{fig-scatters}
\end{figure*}

In the second experiment we fix the number of basis/trees to 100, or at the point where they start to converge from the previous experiment, and generate predictions from {\sc annz2} using the recommended randomized regression script. The number of learning models is set to 100 using both ANNs and Boosted Decision Trees; the remaining options are set to their default values as published. The performance measures are reported in \autoref{table-metrics}, and for the general case of FR$_{e}$, we show in \autoref{fig-FRe} the FR$_{e}$ score as we vary the value of the threshold $e$. The scatter plots for each method are colour coded by the predictive variance and shown in \autoref{fig-scatters}. We find that {\sc gpvc} consistently outperforms all other {\sc gp} methods and also {\sc annz2}, although the margins are at the $\sim 1$ per cent level. {\sc tpz} provides the poorest results by a significant margin for low values of $e$, but asymptotes towards the FR values for the other codes at $e > 0.1$.

 \begin{table}
\caption{Performance measures for each algorithm trained using 100 basis functions for {\sc gpgl}, {\sc gpvl}, {\sc gpvc} and {\sc spgp}, 100 trees for {\sc tpz} and 100 MLMs for {\sc annz2} on the held out test set. The best-performing algorithm is highlighted in bold font.}
\begin{center}
\begin{tabular}{| l | l | l | l | l | l |}
     				&	RMSE		&	MLL		&	FR$_{0.15}$ 		&	FR$_{0.05}$ 	\\	\hline
	{\sc tpz}			&	0.0628		&	1.21		&	98.69\%			&	68.47\%\\
	 {\sc annz2}		&	0.0422		&	1.65		&	98.77\%			&	89.08\%\\
	{\sc spgp}		&	0.0427		&	1.95		&	98.67\%			&	90.60\%\\
	{\sc gpgl}		&	0.0413		&	1.91		&	98.75\%			&	90.45\%\\
	{\sc gpvl}		&	0.0406		&	1.91		&	98.79\%			&	90.16\%\\
	{\sc gpvc}		&	\textbf{0.0388}	&	\textbf{2.00}	&	\textbf{98.90\%}		&	\textbf{91.85\%}\\
  \end{tabular}
\end{center}
\label{table-metrics}
\end{table}

\subsection{Rejection performance}
As stated in \autoref{sec-introduction} one of the critical aspects of using photometric redshifts in future cosmology experiments requires the understanding of the variance on the individual galaxy photometric redshift and on the distribution.

\begin{figure*}
        \centering
        \begin{subfigure}[b]{0.45\textwidth}
                 \includegraphics[width=\textwidth]{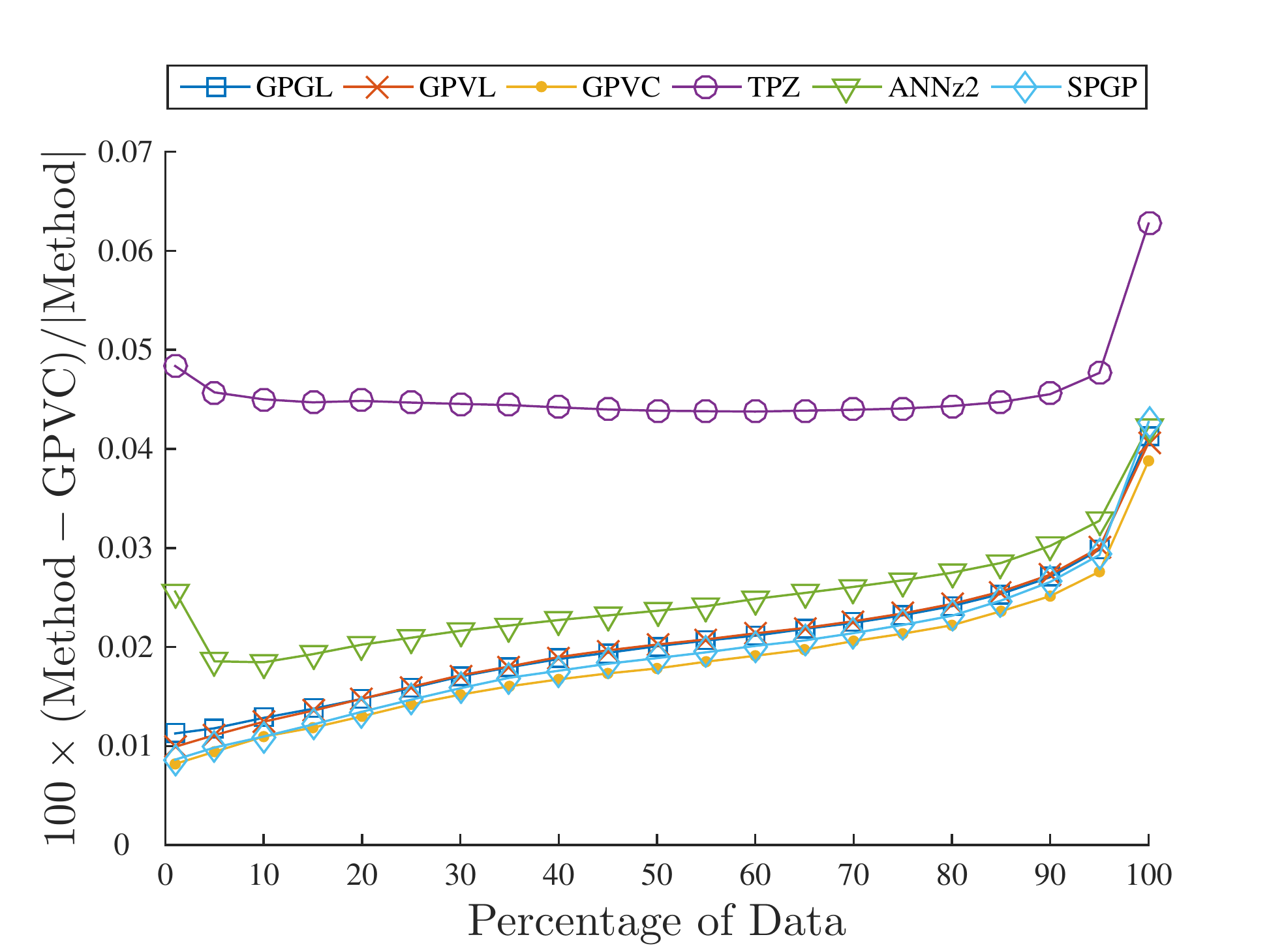}
                 \caption{RMSE}
        \end{subfigure}
        ~
        \begin{subfigure}[b]{0.45\textwidth}
                 \includegraphics[width=\textwidth]{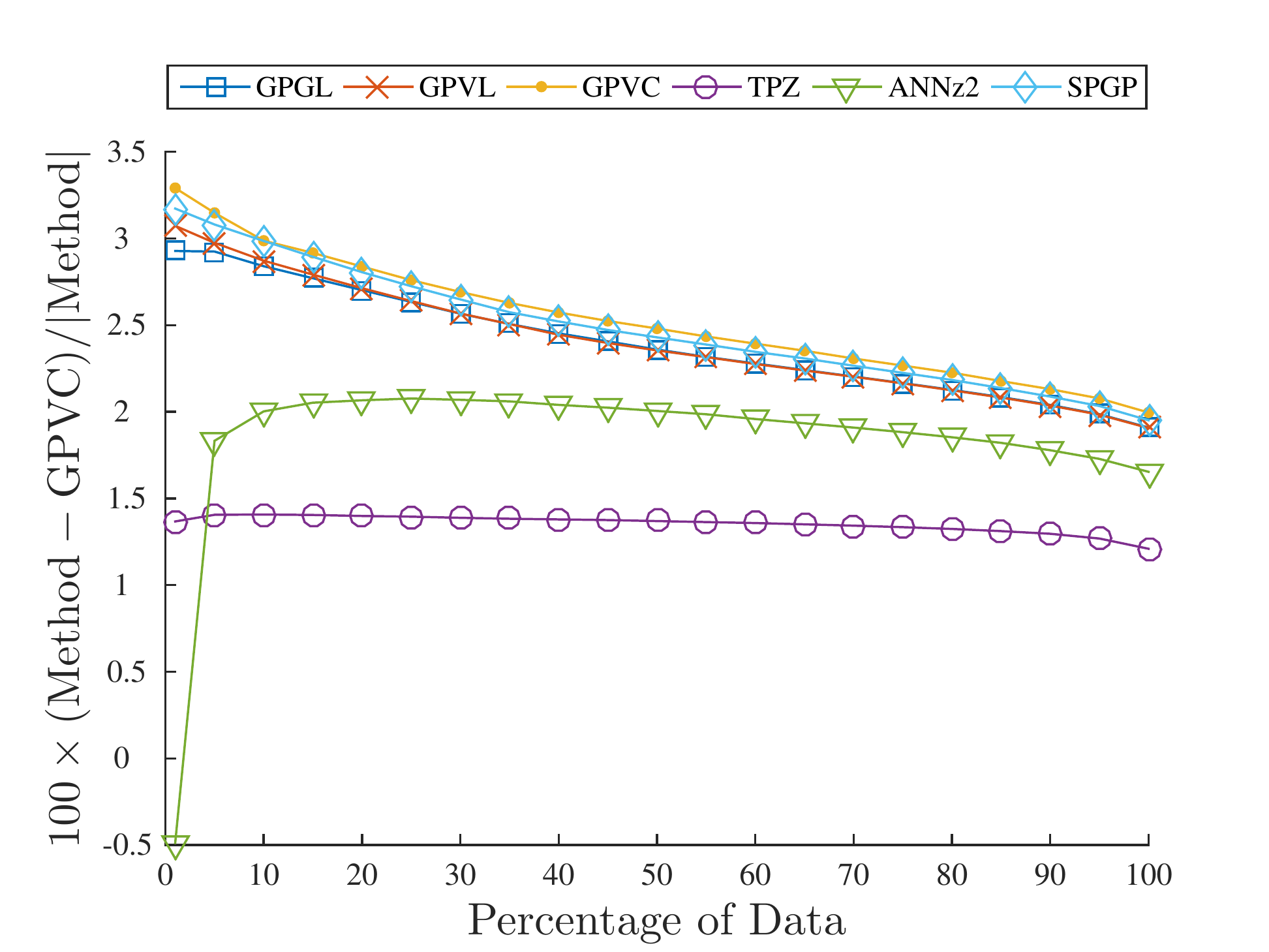}
                 \caption{MLL}
        \end{subfigure}
        
        \begin{subfigure}[b]{0.45\textwidth}
                 \includegraphics[width=\textwidth]{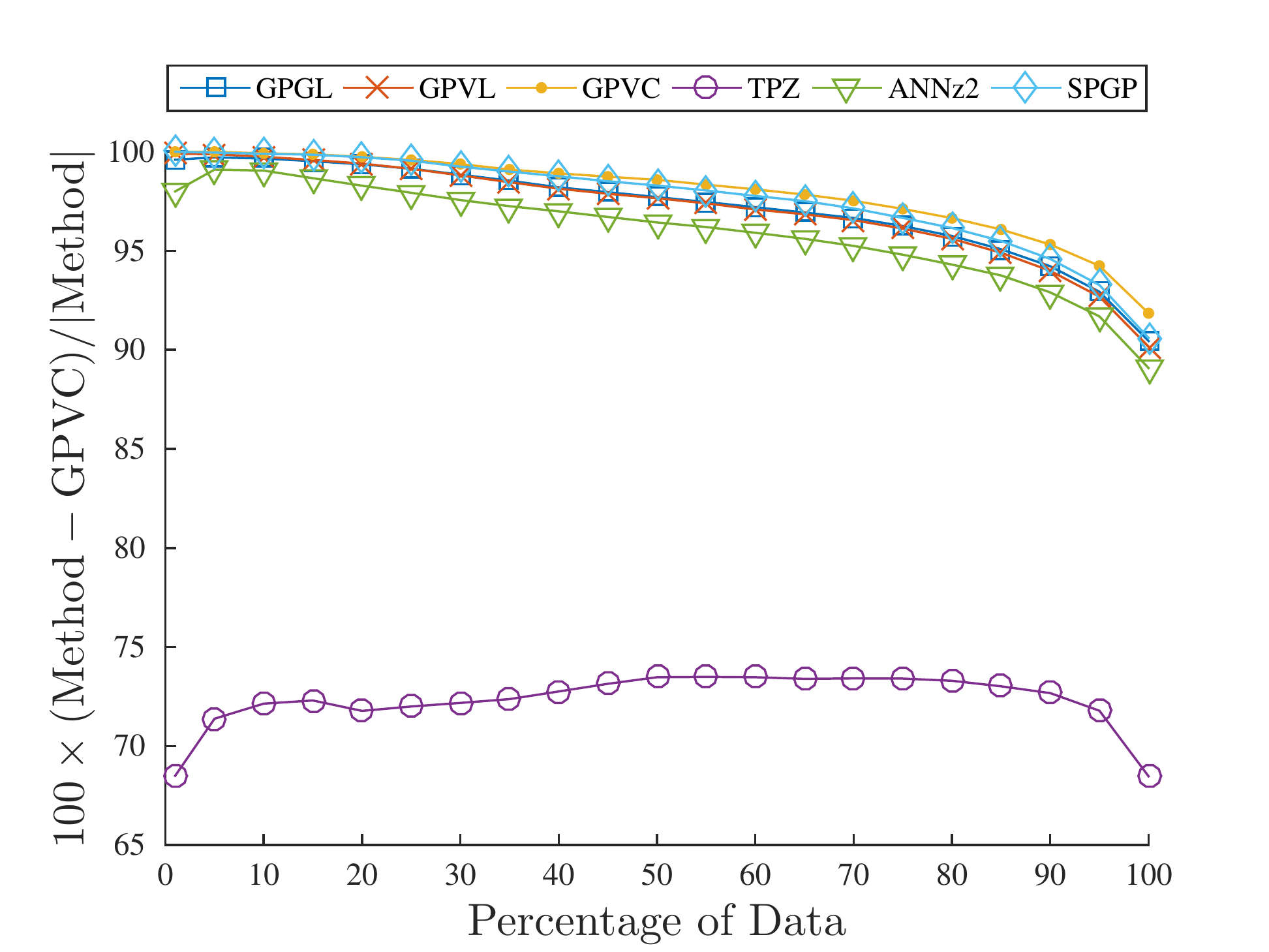}
                 \caption{FR$_{0.05}$}
        \end{subfigure}
        ~
        \begin{subfigure}[b]{0.45\textwidth}
                 \includegraphics[width=\textwidth]{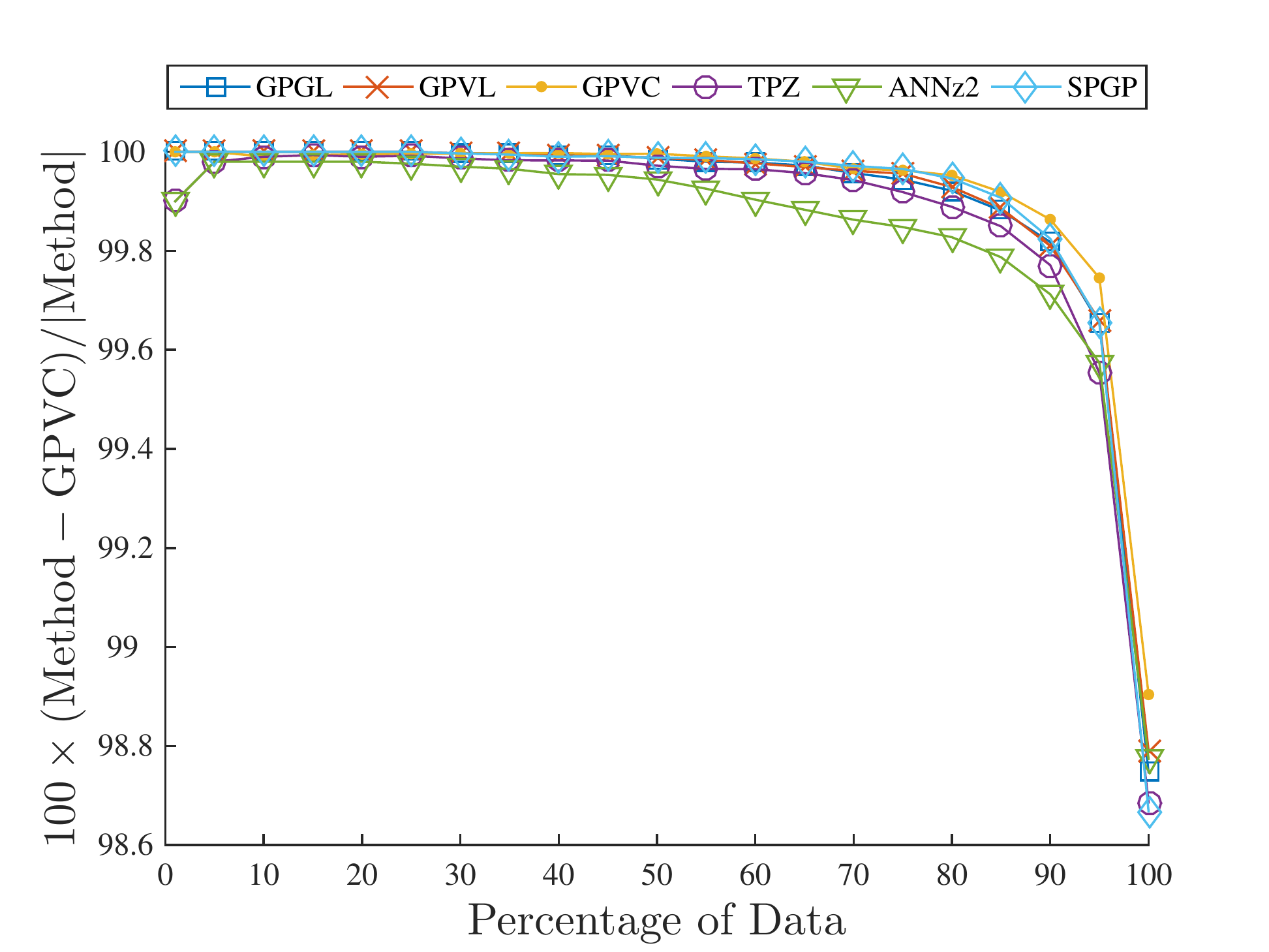}
                 \caption{FR$_{0.15}$}
        \end{subfigure}
        
       \caption{The (a) RMSE, (b) MLL, (c) FR$_{0.05}$ and (d) FR$_{0.15}$ as a function of the percentage of data selected based on the predictive variance generated by each method using 100 basis/trees/MLMs.}
	\label{fig-rejection}
\end{figure*}

\begin{figure*}
        \centering
        \begin{subfigure}[b]{0.45\textwidth}
                 \includegraphics[width=\textwidth]{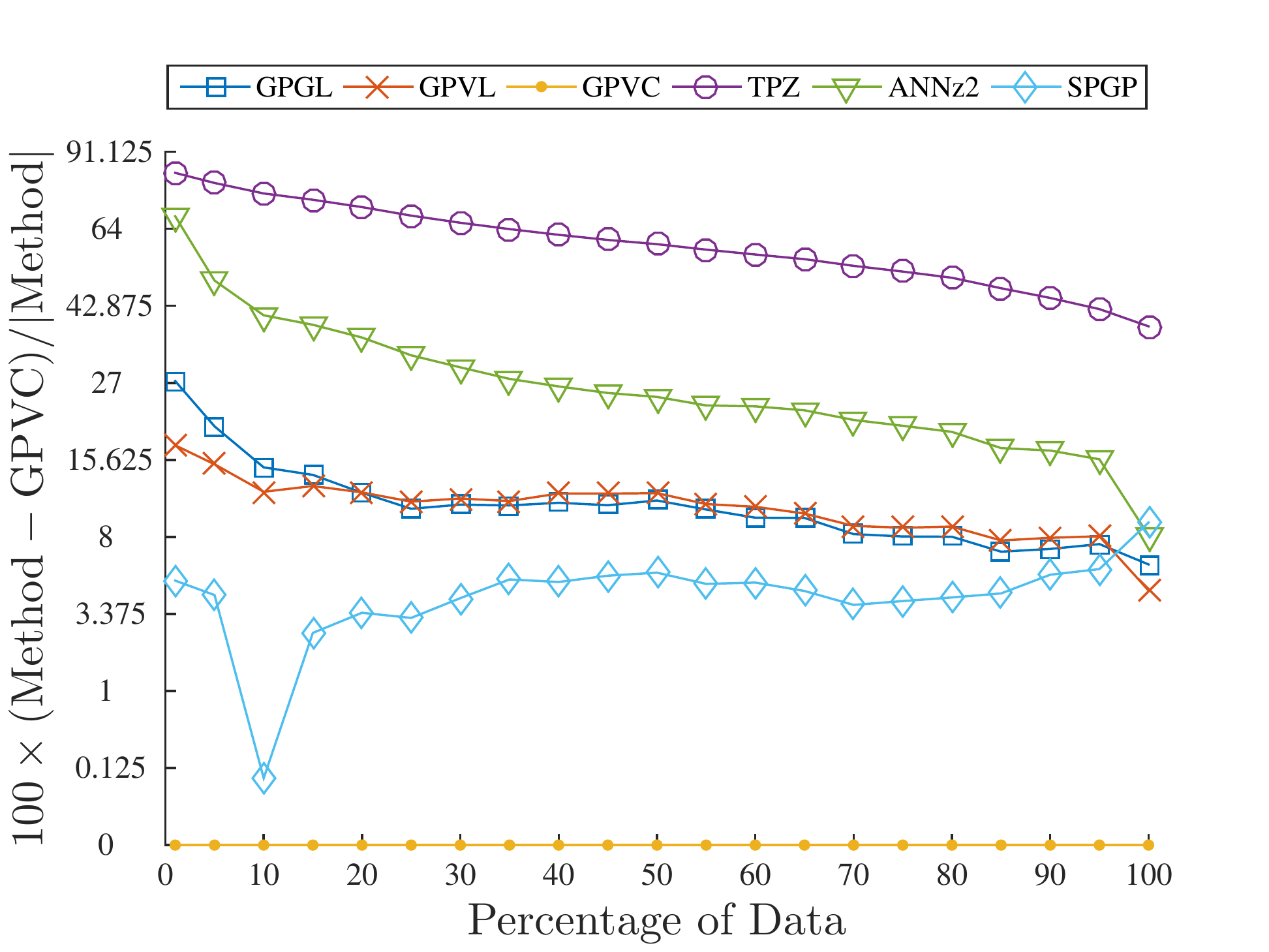}
                 \caption{RMSE}
        \end{subfigure}
        ~
        \begin{subfigure}[b]{0.45\textwidth}
                 \includegraphics[width=\textwidth]{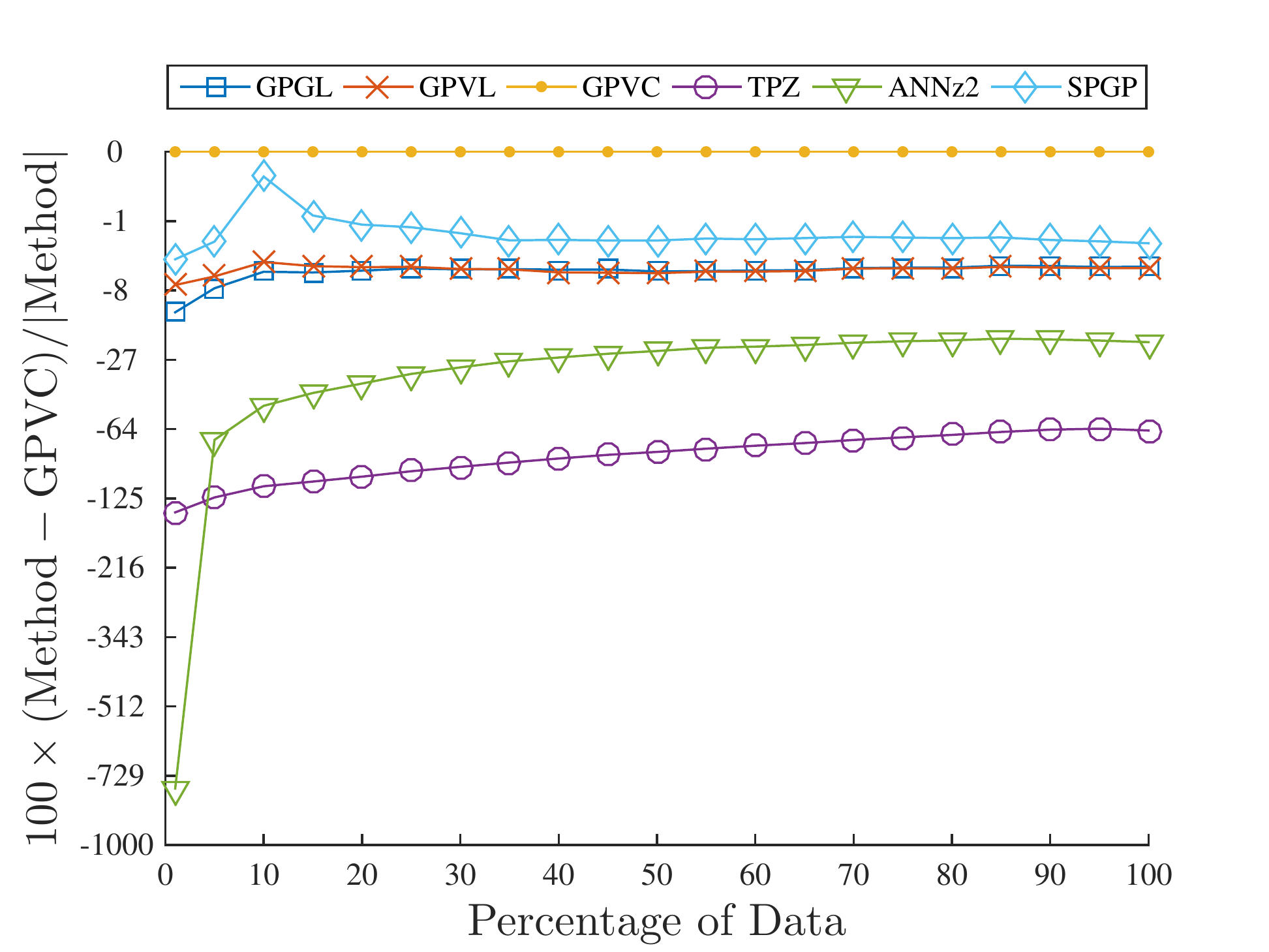}
                 \caption{MLL}
        \end{subfigure}
        
        \begin{subfigure}[b]{0.45\textwidth}
                 \includegraphics[width=\textwidth]{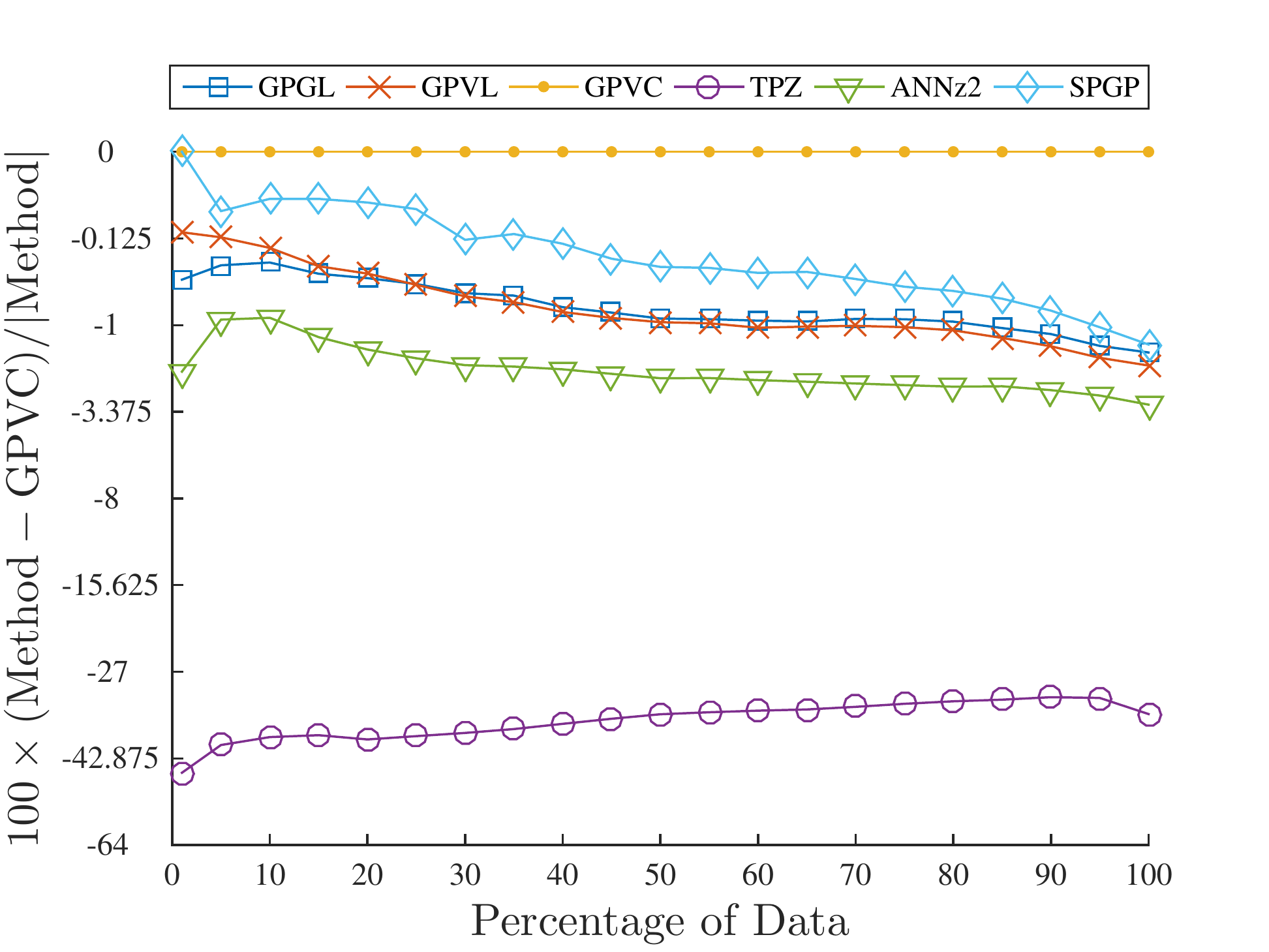}
                 \caption{FR$_{0.05}$}
        \end{subfigure}
        ~
        \begin{subfigure}[b]{0.45\textwidth}
                 \includegraphics[width=\textwidth]{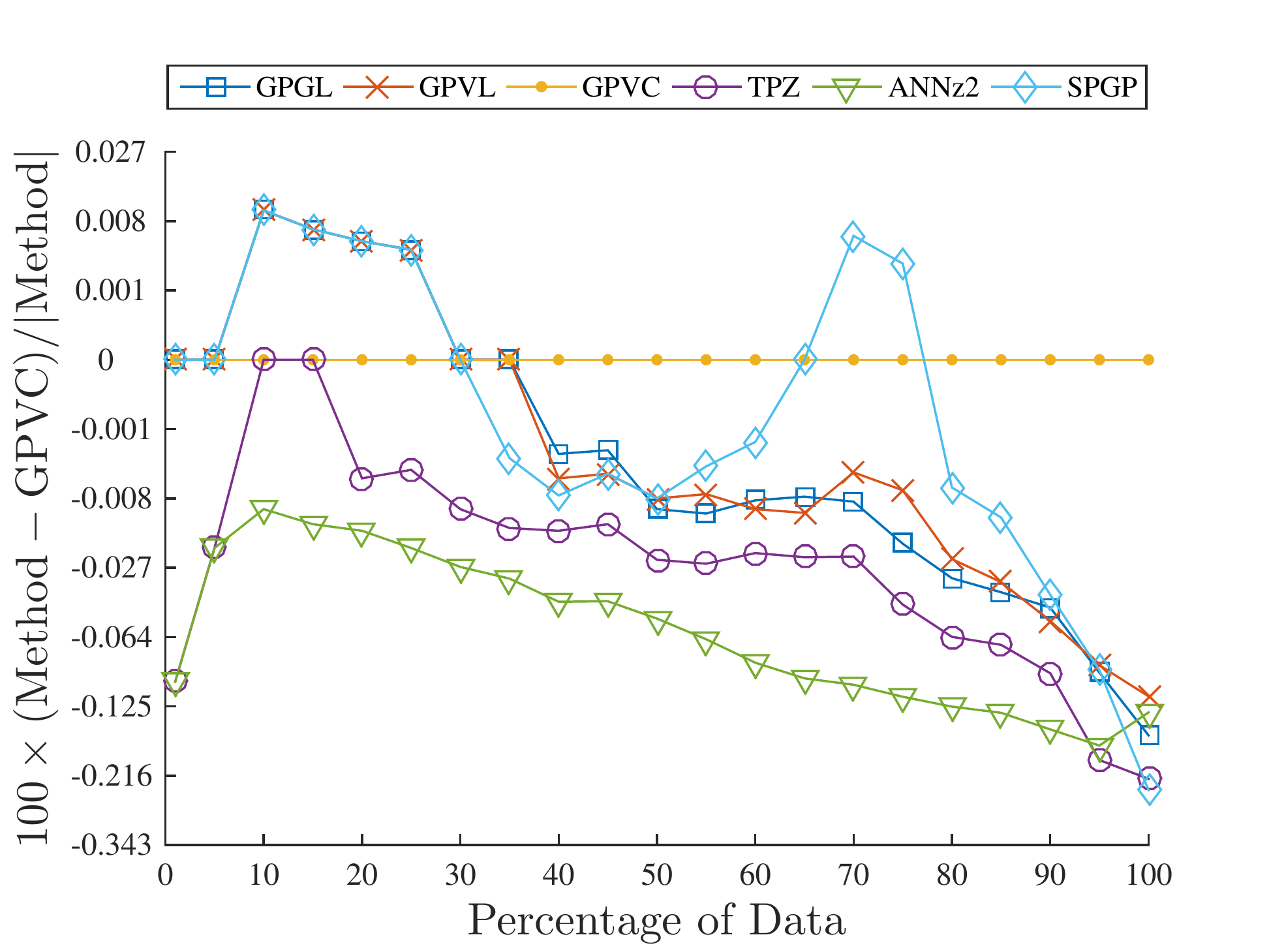}
                 \caption{FR$_{0.15}$}
        \end{subfigure}
        
       \caption{The percentage of difference between {\sc gpvc} and the other methods, computed as $100\times\left(\mbox{Method}-\mbox{GPVC}\right)/\left|\mbox{Method}\right|$, on (a) RMSE, (b) MLL, (c) FR$_{0.05}$ and (d) FR$_{0.15}$ as a function of the percentage of data selected based on the predictive variance generated by each method using 100 basis/trees/MLMs. The values are plotted on a cubic root $y$-axis to enhance visibility.}
	\label{fig-improvement}
\end{figure*}

In this section, we analyse the quality of the models' uncertainty predictions by evaluating their rejection performance, namely their ability to infer which data are associated with high uncertainty; as we remove such samples, we would expect performance to improve. \autoref{fig-rejection} shows the scores of the metrics as a function of the percentage of data selected based on the predictive variance generated by each method using 100 basis/trees/MLMs. {\sc tpz} is significantly worse than the other methods on all metrics, {\sc annz2} performs much better but still underperforms the GP-based methods. {\sc gpgl} and {\sc gpvl} perform equally well, but underperform {\sc spgp} slightly. {\sc gpvc} consistently outperforms the other methods, on all metrics, for almost the entire range. \autoref{fig-improvement} shows the relative change over {\sc gpvc} as a reference for the plots in \autoref{fig-rejection}. {\sc gpvc} shows a significant and consistent improvement over all methods, especially past 20 per cent of the data. For less than 20 per cent, {\sc spgp} is competitive to {\sc gpvc} but is not consistently better. To quantify this, we compute the average improvement that {\sc gpvc} has over the other methods over the entire range; these results are reported in \autoref{table-improvement}. The results show that {\sc gpvc} provides performance improvement over all the other methods on all metrics.
This therefore provides a robust basis for optimizing the sample selection of galaxies to use in various experiments, allowing the trade-off between number of galaxies included and their photometric-redshift accuracy.

\begin{table}
\caption{The average relative improvement of {\sc gpvc} over other tested methods on all metrics on the test set using 100 basis/trees/MLMs.}
\begin{center}
\begin{tabular}{| l | r | r | r | r | r |}
     				&	RMSE		&	MLL		&	FR$_{0.15}$ 		&	FR$_{0.05}$ 	\\	\hline
	{\sc tpz}		&	59.87\%		&	85.76\%	&	0.0448\%			&	35.36\%\\
	 {\sc annz2}	&	27.44\%		&	58.04\%	&	0.0715\%			&	2.03\%\\
	{\sc spgp}		&	4.29\%		&	1.91\%	&	0.0099\%			&	0.326\%\\
	{\sc gpgl}		&	10.89\%		&	5.69\%	&	0.0149\%			&	0.772\%\\
	{\sc gpvl}		&	10.80\%		&	5.02\%	&	0.0137\%			&	0.840\%\\
  \end{tabular}
\end{center}
\label{table-improvement}
\end{table}

\subsection{Bias}
\begin{figure*}
        \centering
        \begin{subfigure}[b]{0.45\textwidth}
                 \includegraphics[width=\textwidth]{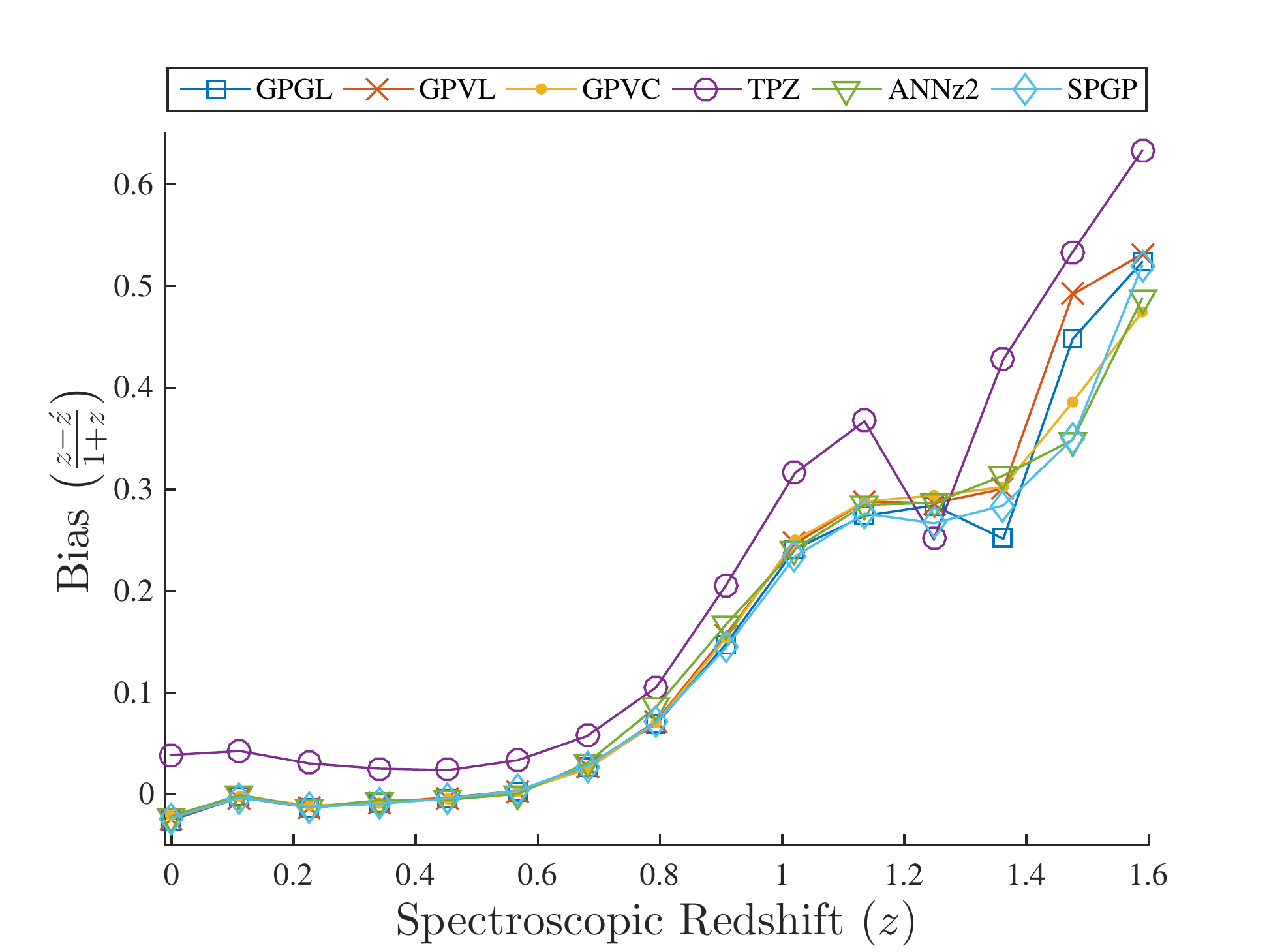}
                 \caption{100\%}
        \end{subfigure}
        ~
        \begin{subfigure}[b]{0.45\textwidth}
                 \includegraphics[width=\textwidth]{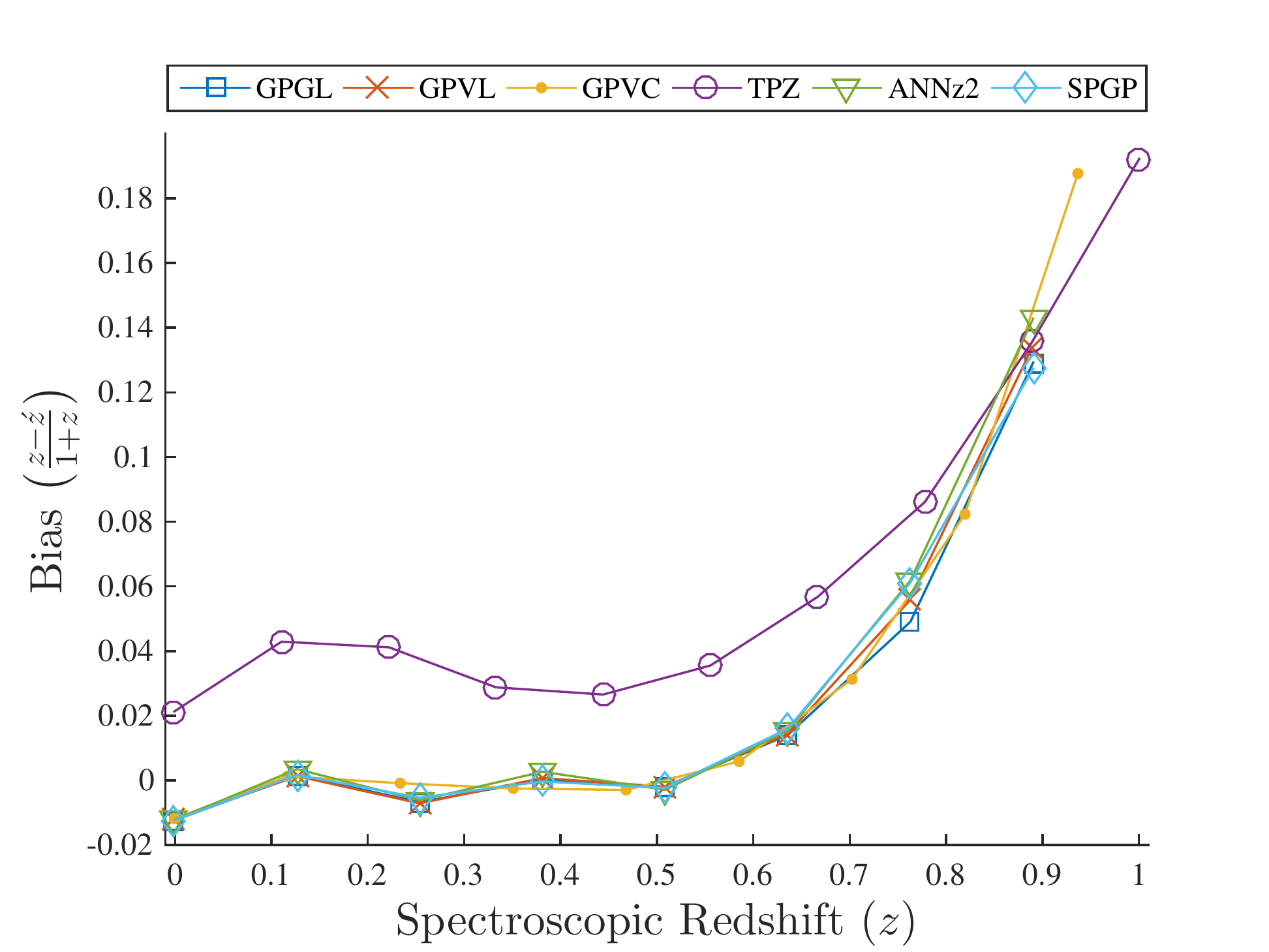}
                 \caption{75\%}
        \end{subfigure}
        
        \begin{subfigure}[b]{0.45\textwidth}
                 \includegraphics[width=\textwidth]{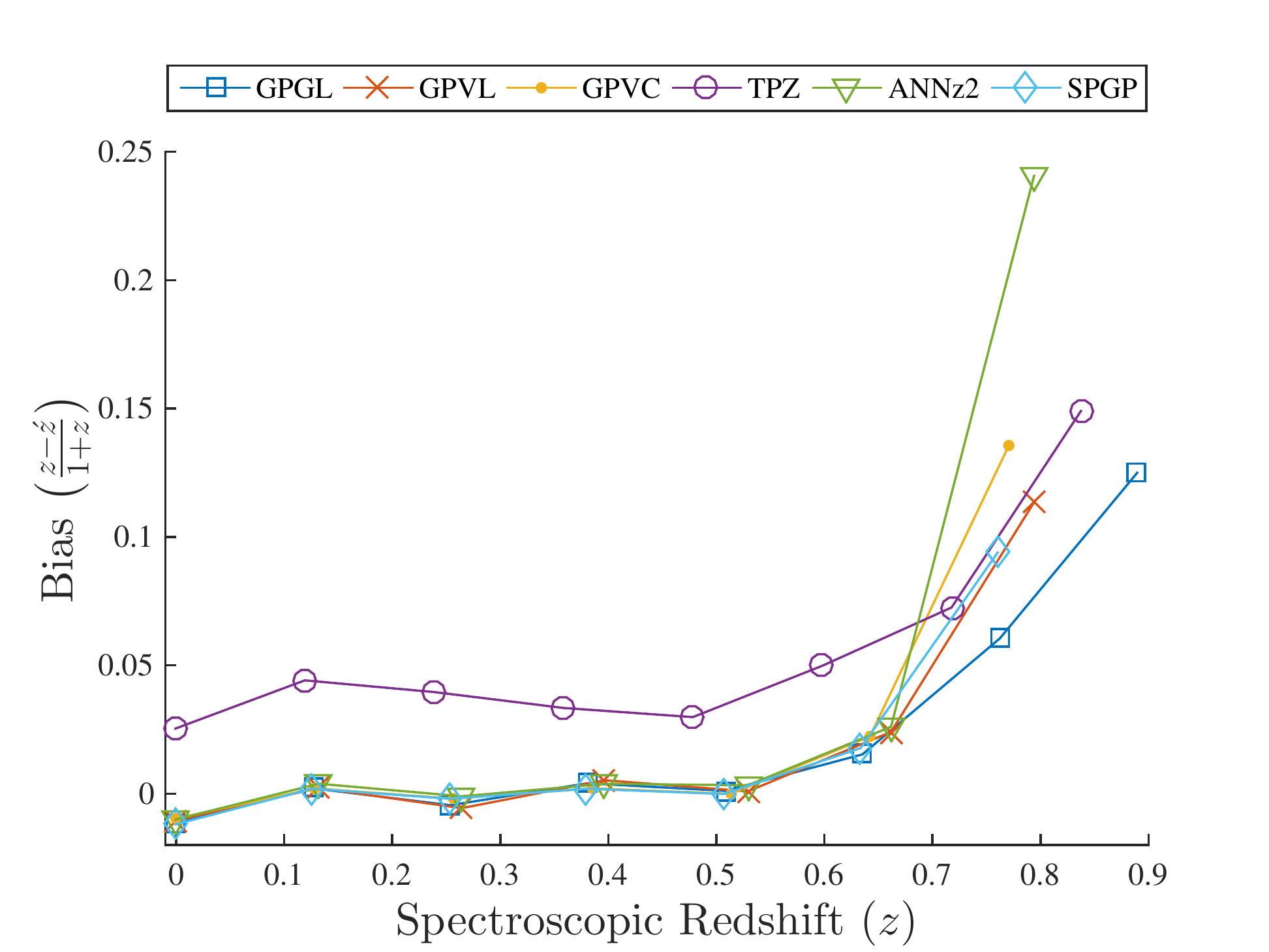}
                 \caption{50\%}
        \end{subfigure}
        ~
        \begin{subfigure}[b]{0.45\textwidth}
                 \includegraphics[width=\textwidth]{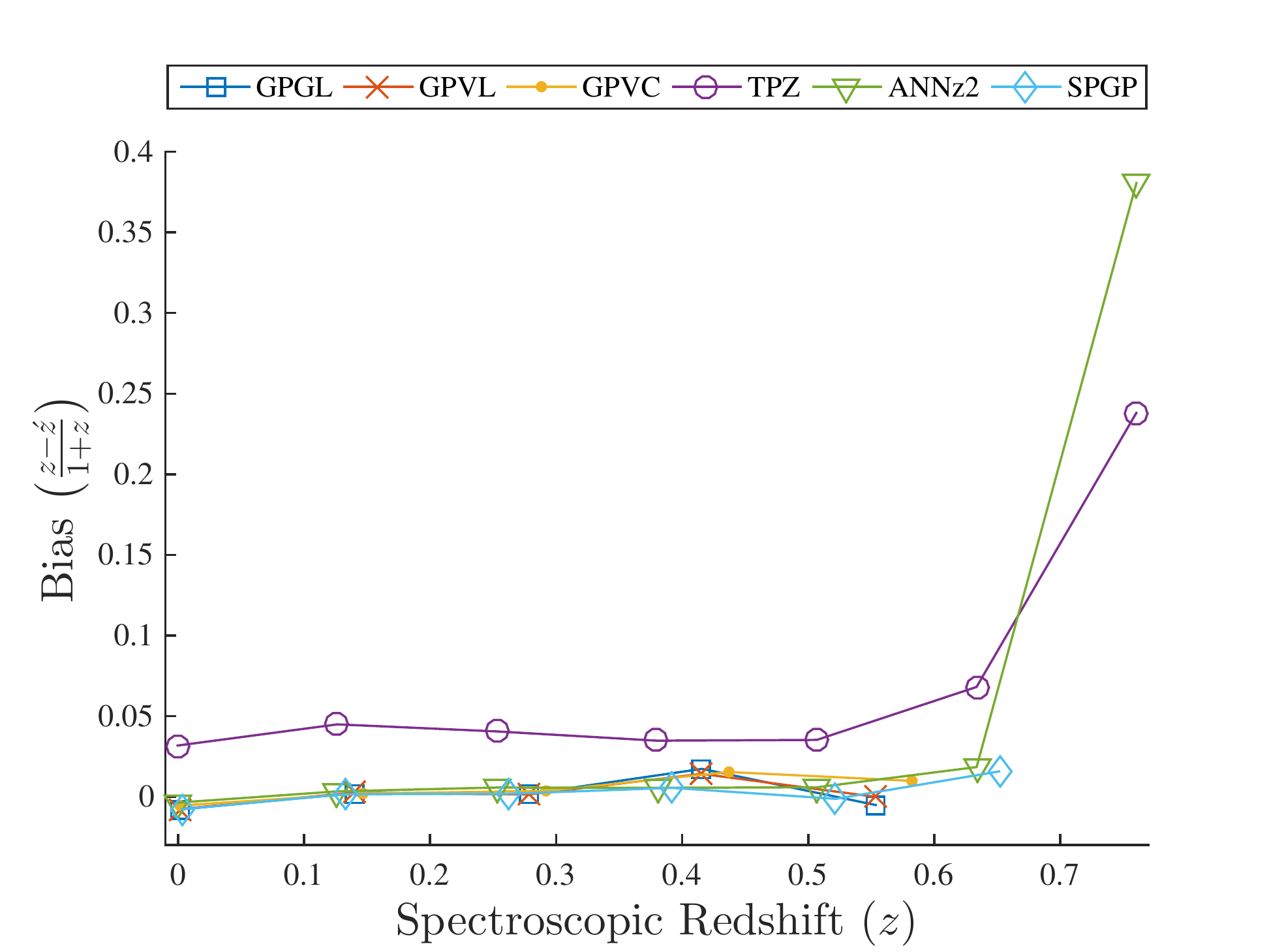}
                 \caption{25\%}
        \end{subfigure}
        
        \caption{The RMSE as a function of the percentage of data selected based on the predictive variance generated by each method using 100 basis/trees/MLMs.}
	\label{fig-bias}
\end{figure*}

We regard bias as a key metric for future experiments and science focus. The bias indicates how the photometric redshift systematically deviates from the true redshift as a function of the input and output. We report in \autoref{fig-bias} the bias (\autoref{eq-bias}) as a function of the spectroscopic redshift ($z$), using different percentages of the data selected by each method's predictive variance grouped by uniformly spaced bins of width 0.1. Over the entire data range, {\sc tpz} shows the worst performance while the remaining methods perform equally well to a redshift of $\sim$ 0.9. At higher redshifts, the performance of the GP-based methods and {\sc annz2} vary with no clear winner. The figure shows that as we exclude more samples, all methods tend to be more certain about low redshift ($z<0.6$) samples. The methods we propose in this paper, however, are more stable and tend to improve as we reject more data, whereas the bias scores for {\sc tpz}, {\sc annz2} and {\sc spgp} in some cases degrade especially for high redshift.

\subsection{Uncertainty Analysis}
\begin{figure*}
        \centering
        \begin{subfigure}[b]{0.45\textwidth}
                 \includegraphics[width=\textwidth]{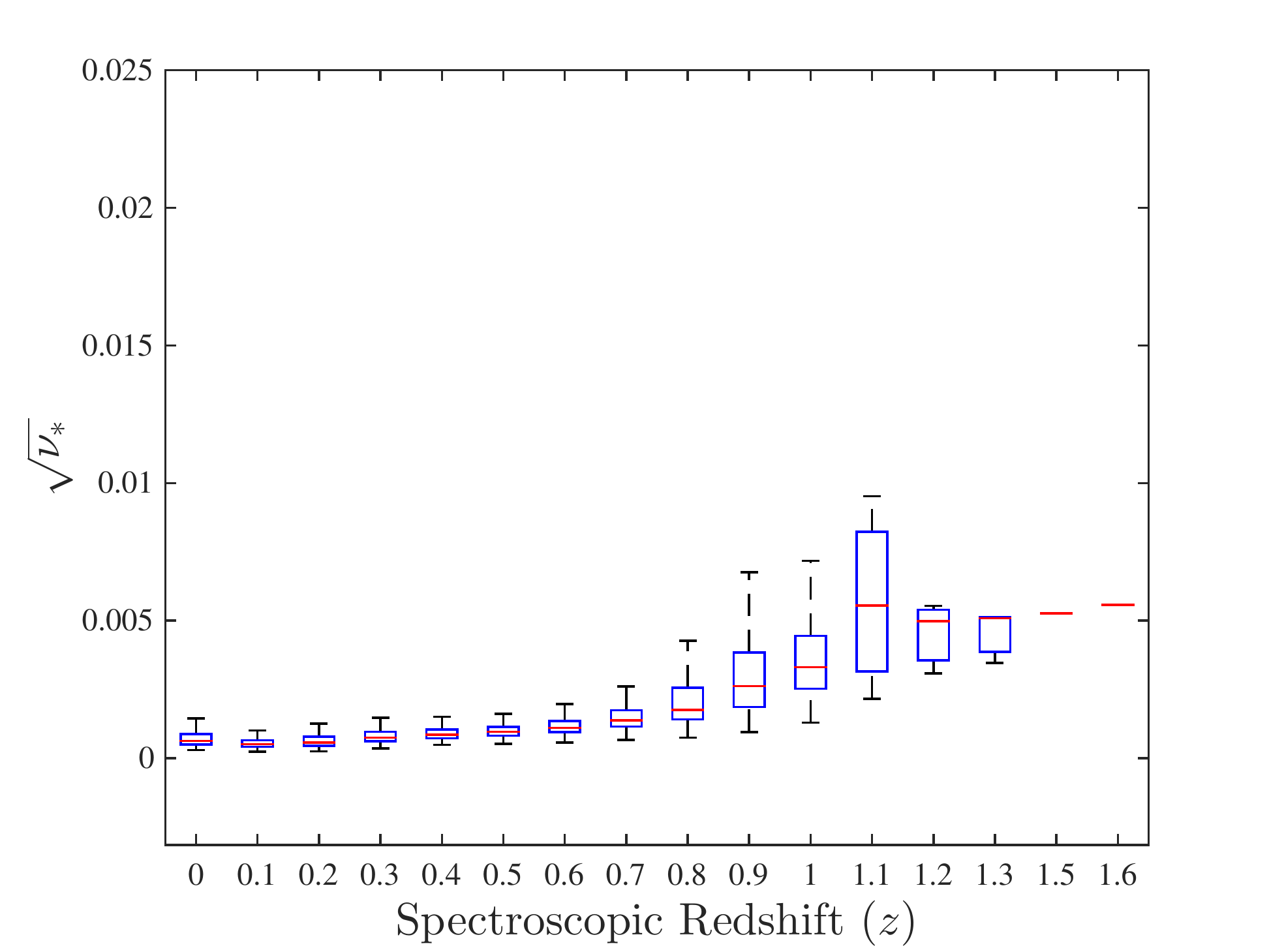}
                 \caption{Model uncertainty.}
        \end{subfigure}
        ~
        \begin{subfigure}[b]{0.45\textwidth}
                 \includegraphics[width=\textwidth]{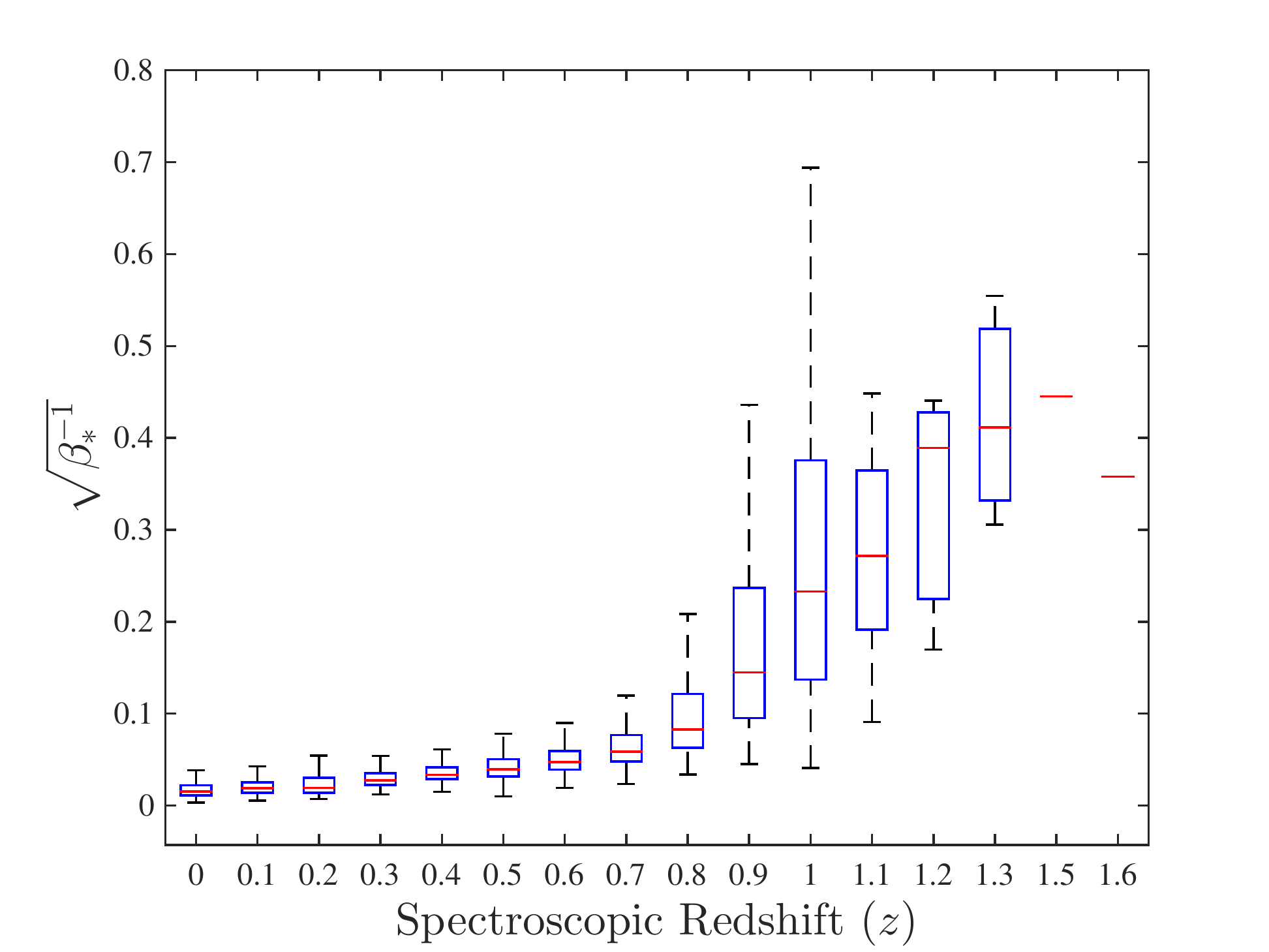}
                 \caption{Noise uncertainty.}
        \end{subfigure}
        
       \caption{A box plots of the square root of (a) the model uncertainty (\autoref{eq-nu}) and the (b) noise uncertainty (\autoref{eq-beta}), produced by a {\sc gpvc} model with 100 basis functions, vs. the spectroscopic redshift showing median (bar), inter-quartile range (box) and range (whiskers).}
	\label{fig-model-noise}
\end{figure*}

As discussed in \autoref{sec-heteroscedastic-noise}, the predictive variance produced by the proposed {\sc gpvc} method is composed of two terms that model the uncertainty about the function due to data density inhomogeneity and the noise uncertainty. In this experiment, we analyse these two components of uncertainty separately using a {\sc gpvc} model with 100 basis functions. \autoref{fig-model-noise} shows the model and noise uncertainties as functions of the spectroscopic redshift ($z$) using uniformly spaced bins of width 0.1. Both start to increase rapidly beyond $z \sim 0.5$. However, the overwhelming contribution to the overall uncertainty for high redshifts is due to the intrinsic noise rather than the scarcity of data. This indicates that the amount of data is sufficient for the model to be confident about its mean function and we have precise enough features for redshifts $<0.5$. For higher redshifts, the results indicate that obtaining more precise, or additional, features (e.g. near-infrared photometry) is a better investment than obtaining, or training on, more samples. This is not a surprising result given the data used, i.e. the spectroscopic training set and the test set are both sub-samples derived from the same overall SDSS galaxy sample. However, such a situation will not be the case for most cosmological applications that require photometric redshifts, and having such separable noise terms will aid in determining the optimal approach to ensure that the requisite training samples are in place to address particular scientific problems, from galaxy evolution to various cosmology experiments. 

\subsection{Time complexity analysis}
We provide in this section analysis of the theoretical and empirical time complexity of the methods tested in this paper. \autoref{table-time-complexity} shows the upper bound time complexity of each algorithm as a function of the number of input data samples $n$, features $d$ (the dimensionality of the input data) and basis functions/trees/neurons $m$. If $m\ge d^{2}$, then the time complexity of {\sc gpgl}, {\sc gpvl}, {\sc gpvc} and {\sc spgp} is equal to $O\left(nm^{2}\right)$, whereas the time complexities of {\sc tpz} and a single layer neural network will remain the same. Thus, for $m<d\log\left(n\right)$ random forest trees have a higher upper bound than the other methods. The time complexity of random forests can be reduced to $O\left(n_{s}md_{s}\mathcal{D}\right)$, where $n_{s}$ is the subsample size to grow each tree, $d_{s}$ is the number of subsampled features used to grow each tree and $\mathcal{D}$ is the maximum depth allowed for each tree to grow. In practice however, efficient implementation of the methods can significantly impact the actual running time. For example, effective computing of matrix operations ideally makes use of algorithms that are parallelizable and hence can be even further accelerated using graphical processing units (GPUs). Using the same training data set of 100,000 samples, it took {\sc tpz} 1 hour, 46 minutes and 42 seconds to train a forest of 5 trees; whereas {\sc annz2}'s random forest implementation required only 4 minutes and 34 seconds. On the other hand, the random forest implementation of {\sc matlab}'s statistical and machine learning toolbox, and {\sc python}'s {\sc sklearn} library required less than 3 seconds. Training a single layer neural network with 5 hidden neurons using {\sc annz2} for 500 iterations required 20 minutes and 41 seconds, whereas {\sc spgp}, {\sc gpgl}, {\sc gpvl} and {\sc gpvc} trained using 5 basis functions for the same number of iterations required 2 minutes and 5 seconds, 1 minute and 28 seconds, 1 minute and 28 seconds, and 2 minutes and 32 seconds respectively.

\begin{table}
\caption{The theoretical time complexity of each approach, where $n$ is the number of samples, $d$ is the number of features or the dimensionality of the input, $m$ is the number of basis functions, trees in TPZ and hidden units in a single layer ANN}
\begin{center}
  \begin{tabular}{| l | l |}
     	Method		&	Time complexity					\\	\hline				\\
	{\sc ANN} 		&	$O\left(nmd\right)$					\\
	{\sc tpz} 		&	$O\left(nmd\log\left(n\right)\right)$					\\
	{\sc spgp}		&	$O\left(nmd+nm^{2}\right)$				\\
	{\sc gpgl}, {\sc gpvl}, {\sc gpgd} and {\sc gpvd}		&	$O\left(nmd+nm^{2}\right)$		\\	
	{\sc gpgc} and {\sc gpvc}		&	$O\left(nmd^{2}+nm^{2}\right)$	\\	\hline
  \end{tabular}
\end{center}
\label{table-time-complexity}
\end{table}

\section{Conclusions}
\label{sec-conclusion}
We have produced and implemented an extension of the sparse Gaussian process framework presented in \cite{almosallam2015} to incorporate separable terms for intrinsic noise in the data and the model uncertainty due to the finite data samples in the training set. These are combined to estimate the total variance on the predicted photometric redshifts. 

We find that our algorithm outperforms other machine learning methods tested in the literate across all metrics considered. In particular, we find that by including these terms we are able to accurately determine the relative variance between photometric redshift in individual galaxies. This leads to the ability to reject parts of the data set in order to gain higher accuracy on the required metric, i.e. root-mean square error, normalized median absolute deviation and/or the bias as a function of redshift. Moreover, the presented models provide a significant time improvement especially over {\sc tpz} and {\sc annz2}. The algorithm, which includes the cost-sensitive learning discussed in \cite{almosallam2015}, in addition to the separable noise terms presented in this paper is available in {\sc matlab} and {\sc python} implementations from \url{https://github.com/OxfordML/GPz}.

In a subsequent paper, we will investigate how the algorithm can be used to define future imaging and spectroscopic surveys in order to provide the most efficient strategy for delivering photometric redshifts of the accuracy required to perform various cosmological experiments with future facilities, similar to the work of \cite{Masters2015} but with the added advantage of being able to separate data density issues from uncertainty due to photometric noise.

\section*{Acknowledgments}
IAA acknowledges the support of King Abdulaziz City for Science and Technology.
MJJ acknowledges support from the UK Space Agency.

\bibliographystyle{mnras}
\bibliography{photoz-var}

\appendix

\section{Gaussian Processes}
\label{append-gaussian-processes}
A Gaussian Process is a supervised non-linear regression algorithm lying within the class of Bayesian non-parametric models due to the few explicit \emph{parametric} assumptions that it makes about the nature of the function fit. Given a set of input $\mathbfss{X}=\left\{\mathbfit{x}_{i}\right\}_{i=1}^{n}\in \mathbb{R}^{n\times d}$ and a set of target outputs $\mathbfit{y}=\left\{y_{i}\right\}_{i=1}^{n}\in \mathbb{R}^{n}$, where $n$ is the number of samples in the dataset and $d$ is the dimensionality of the input, the underlying assumption of a GP is that the observed target $y_{i}$ is generated by a function of the input $\mathbfit{x}_{i}$ plus additive noise $\epsilon_{i}$:
\begin{align}
y_{i} = f\left(\mathbfit{x}_{i}\right)+\epsilon_{i},
\end{align}
where $\epsilon\sim\mathcal{N} \left(0,\sigma^{2}\right)$. It is assumed that $\mathbfit{y}$ has a zero mean (this can readily be achieved without loss of generality) and univariate, although the derivation can be readily extended to the multivariable case. The likelihood, the probability of observing the targets given the function, is hence distributed as follows:
\begin{align}
p\left(\mathbfit{y}|\mathbfit{f}_{x},\sigma^{2}\right)=\mathcal{N} \left(\mathbfit{f}_{x},\sigma^{2}\mathbfss{I}\right),
\end{align}
where $\mathbfit{f}_{x}=\{f\left(\mathbfit{x}_{1}\right),\hdots,f\left(\mathbfit{x}_{n}\right)\}$. A GP then proceeds by applying Bayes theorem to infer the sought after distribution of the function $\mathbfit{f}_{x}$ given the observations:
\begin{align}
p\left(\mathbfit{f}_{x}|\mathbfit{y},\mathbfss{X},\sigma^{2}\right) = \frac{p\left(\mathbfit{y}|\mathbfit{f}_{x},\sigma^{2}\right)p\left(\mathbfit{f}_{x}|\mathbfss{X}\right)}{p\left(\mathbfit{y}|\mathbfss{X},\sigma^{2}\right)}.
\end{align}
This requires us to define a prior, $p\left(\mathbfit{f}_{x}|\mathbfss{X}\right)$, over our space of functions. Most widely used priors assume local similarity in the data, i.e. closeby inputs are mapped to similar outputs. More formally, we assume a normally distributed prior with a mean of zero, to match the mean of the normalized target $\mathbfit{y}$, with a covariance \emph{function} $\mathbfss{K}$ to capture our prior belief of data locality, i.e. $p\left(\mathbfit{f}_{x}|\mathbfss{X}\right)\sim\mathcal{N} \left(0,\mathbfss{K}\right)$. The covariance $\mathbfss{K}$ is modelled as a function of the input, $\mathbfss{K}=\kappa\left(\mathbfss{X},\mathbfss{X}\right)$. Each element at the $i$-th row and the $j$-th column of $\mathbfss{K}$ is set equal to $\kappa\left(\mathbfit{x}_{i},\mathbfit{x}_{j}\right)$, where $\kappa$ is the covariance function. The function $\kappa$ cannot be any arbitrary mapping, as it has to guarantee that $\mathbfss{K}$ is a valid covariance matrix, i.e. symmetric and positive semi-definite. A class of functions referred to as  \emph{Mercer kernels} guarantees these structural constraints \citep{mercer1909}. An example of valid Mercer kernel is the squared exponential kernel, defined as follows:
\begin{align}
\kappa\left(\mathbfit{x}_{i},\mathbfit{x}_{j}\right) = h^{2}\exp\left(-\frac{1}{2\lambda^{2}}\left\|\mathbfit{x}_{i}-\mathbfit{x}_{j}\right\|^{2}\right),
\label{eq-squared-exponential}
\end{align}
where $h$ and $\lambda$ are referred to as the height and length scale respectively. The reader is referred to \cite{rasmussen2006gaussian} or \cite{roberts2012rs} for in depth discussion of covariance functions and kernels. With a likelihood $p\left(\mathbfit{y}|\mathbfit{f}_{x}\right)$ and a prior $p\left(\mathbfit{f}_{x}|\mathbfss{X}\right)$, the marginal likelihood $p\left(\mathbfit{y}|\mathbfss{X}\right)$ can be computed as follows \cite{rasmussen2006gaussian}:
\begin{align}
p\left(\mathbfit{y}|\mathbfss{X},\sigma^{2}\right) =& \int p\left(\mathbfit{y}|\mathbfit{f}_{x},\mathbfss{X},\sigma^{2}\right)p\left(\mathbfit{f}_{x}|\mathbfss{X}\right) \mathrm{d}\mathbfit{f}_{x}\label{eq-marginal-likelihood-complete}\\
=& \mathcal{N} \left(0,\mathbfss{K}+\sigma^{2}\mathbfss{I}\right).
\end{align}
The parameters of the kernel and the noise variance, collectively referred to as the \emph{hyper-parameters} of the model, are then optimized by maximizing the probability of the log of the marginal likelihood in \autoref{eq-marginal-likelihood-complete}:
\begin{align}
\ln p(\mathbfit{y}|\mathbfss{X},\sigma^{2}) =& -\frac{1}{2}\mathbfit{y}^{T}\left(\mathbfss{K}+\sigma^{2}\mathbfss{I} \right)^{-1}\mathbfit{y}\label{eq-log-marginal-likelihood}\nonumber \\
&-\frac{1}{2} \ln\left | \mathbfss{K}+\sigma^{2}\mathbfit{I}\right|-\frac{n}{2}\ln(2\pi).
\end{align}
Once the hyper-parameters have been inferred, the probability of future predictions $\mathbfit{f}_{*}$ for test cases $\mathbfss{X}_{*}$ given the training set, the predictive distribution, can be inferred from the joint distribution of $\mathbfit{f}_{*}$ and the observed targets $\mathbfit{y}$. If we assume that the joint distribution is a multivariate Gaussian, then the joint probability is distributed as follows:
\begin{align}
p\left ( \mathbfit{y},\mathbfit{f}_{*}|\mathbfss{X},\mathbfss{X}_{*},\sigma^{2}\right) = \mathcal{N} \left (0, \begin{bmatrix}\mathbfss{K}_{xx}+\sigma^{2}\mathbfss{I} & \mathbfss{K}_{x*}\\\mathbfss{K}_{*x} & \mathbfss{K}_{**} \end{bmatrix}\right ),
\end{align}
where $\mathbfss{K}_{xx}=\kappa\left(\mathbfss{X},\mathbfss{X}\right)$, $\mathbfss{K}_{x*}=\kappa\left(\mathbfss{X},\mathbfss{X}_{*}\right)$, $\mathbfss{K}_{*x}=\kappa\left(\mathbfss{X}_{*},\mathbfss{X}\right)$ and $\mathbfss{K}_{**}=\kappa\left(\mathbfss{X}_{*},\mathbfss{X}_{*}\right)$. The predictive distribution $p\left(\mathbfit{f}_{*}|\mathbfit{y},\mathbfss{X},\mathbfss{X}_{*},\sigma^{2}\right)$ is therefore distributed normal with the following mean and variance:
\begin{align}
\mu_{*} &= \mathbfss{K}_{*x}\left(\mathbfss{K}_{xx}+\sigma^{2}\mathbfss{I}\right)^{-1}\mathbfit{y},\\
\sigma_{*}^{2}	&= \mathbfss{K}_{**}-\mathbfss{K}_{*x}\left(\mathbfss{K}_{xx}+\sigma^{2}\mathbfss{I}\right)^{-1}\mathbfss{K}_{x*}+\sigma^{2}.
\end{align}

\section{The Relation between Sparse GPs and Artificial Neural Networks}
\label{append-relation-to-ann}
An artificial neural networks for regression is a special case of basis function models where the basis functions are sigmoid activations, i.e. $\phi_{j}\left(\mathbfit{x}_{i}\right)=\mbox{sigmoid}\left(\mathbfit{x}_{i}\mathbfit{p}_{j}^{T}+b_{j}\right)$, where $\mathbfit{p}_{j}$ plays the role of the weights between the input an the hidden neuron $j$. The activations of the $m$ hidden units for the $n$ samples in a single-layer ANN is essentially the $\Phi$ matrix. The wight parameters $\mathbfit{w}$ in an ANN regressor are the connections between the hidden units and the output layer. For the neurons' bias terms, they can be simply incorporated by augmenting the input vector and the basis response vector with an additional constant value of 1. Thus, a single layer ANN with $m$ hidden units is a BFM with $m$ basis functions, set as the sigmoid function, and an additional basis function with a constant output of 1. A main distinction between them however, is that the weight parameters $\mathbfit{w}$ in ANNs are treated as parameters of the models to be optimized and are not integerated out. Moreover, the objective function to be optimized in ANNs is different.  Unlike the log marginal likelihood in BFMs, the objective in ANNs is to minimize the regularized sum of squares:

\begin{align}
\mathcal{L}\left(\btheta\right) = \frac{1}{2} \left\|\Phi\mathbfit{w}-\mathbfit{y}\right\|^{2}+\frac{\lambda}{2} \mathbfit{w}^{T}\mathbfit{w}+\frac{\lambda}{2} \sum_{j=1}^{m}\mathbfit{p}_{j}^{T}\mathbfit{p}_{j},
\label{eq-ann-objective}
\end{align}
where $\btheta=\left\{\mathbfit{w},\mathbfit{p}_{1},\hdots\mathbfit{p}_{m},b_{1},\hdots,b_{m}\right\}$ is the set of free parameters to be optimized. We recognize the first two terms as the negative of two terms in the log marginal likelihood defined in \autoref{eq-log-marginal-bfm}, with $\lambda=\alpha/\beta$. Note that unlike the proposed approach where we model each weight with its bespoke precision parameter and each input with its own predictive variance, typical ANNs implicitly assume a constant noise width. Moreover, $\lambda$ is typically treated as an input parameter tuned by cross validation rather than a parameter of the model to be optimized. Another distinction is that ANNs also minimize the norm of the weights in the hidden layer as well as the output layer with no penalty on the bias terms. Moreover, the $\ln\left|\Sigma\right|$ term is missing from \autoref{eq-ann-objective}, which is very crucial as it drives the optimization process towards reducing the uncertainty on the parameter $\mathbfit{w}$, thus producing more confident models with more accurate variance prediction.

\section{Optimization of Sparse Gaussian Processes}
\label{append-optimization}
To ensure that the $\balpha$'s and $\boldeta$'s are positive, we optimize with respect to the log of the parameters. We refer to the set of free parameters to be optimized as $\btheta=\left\{\mathbfss{P},\Gamma_{1},\hdots,\Gamma_{m},\mathbfit{u},b,\ln\balpha,\ln\boldeta\right\}$. The derivative of the log marginal likelihood in \autoref{eq-log-marginal-var} with respect to each parameter $\theta_{i}$ can be found by computing the following in order:

\begin{align}
\frac{\upartial \Sigma}{\upartial \theta_{i}} =& \Phi^{T}\left( \frac{\upartial \mathbfss{B}}{\upartial \theta_{i}}\Phi+2\mathbfss{B} \frac{\upartial \Phi}{\upartial \theta_{i}}\right)+\frac{\upartial \mathbfss{A}}{\upartial \theta_{i}},\\
\frac{\upartial \bar{\mathbfit{w}}}{\upartial \theta_{i}} =& \Sigma^{-1}\left(\Phi^{T}\frac{\upartial \mathbfss{B}}{\upartial \theta_{i}} \mathbfit{y}+\frac{\upartial \Phi^{T}}{\upartial \theta_{i}}\mathbfss{B} \mathbfit{y}-\frac{\upartial \Sigma}{\upartial \theta_{i}}\bar{\mathbfit{w}}\right),\\
\frac{\upartial \bdelta}{\upartial \theta_{i}}=&\frac{\upartial \Phi}{\upartial \theta_{i}}\bar{\mathbfit{w}}+\Phi\frac{\upartial \bar{\mathbfit{w}}}{\upartial \theta_{i}},\\
\frac{\upartial \ln p\left(\mathbfit{y}\right)}{\upartial \theta_{i}} =& -\frac{1}{2}\bdelta^{T} \left(\frac{\upartial \mathbfss{B}}{\upartial \theta_{i}}\bdelta+2\mathbfss{B}\frac{\upartial \bdelta}{\upartial \theta_{i}}\right)\label{eq-dLdt}\\
&-\frac{1}{2}\bar{\mathbfit{w}}^{T}\left(\frac{\upartial \mathbfss{A}}{\upartial \theta_{i}}\bar{\mathbfit{w}}+2\mathbfss{A}\frac{\upartial \bar{\mathbfit{w}}}{\upartial \theta_{i}}\right)\nonumber\\
&-\frac{1}{2}\mathbfit{u}^{T}\left(\frac{\upartial \mathbfss{N}}{\upartial \theta_{i}}\mathbfit{u}+2\mathbfss{N}\frac{\upartial \mathbfit{u}}{\upartial \theta_{i}}\right)\nonumber\\
&+\frac{1}{2}\mbox{trace}\left(\Sigma^{-1}\frac{\upartial \Sigma}{\upartial \theta_{i}}\right)+\frac{1}{2}\mbox{trace}\left(\mathbfss{B}^{-1}\frac{\upartial \mathbfss{B}}{\upartial \theta_{i}}\right)\nonumber\\
&+\frac{1}{2}\mbox{trace}\left(\mathbfss{A}^{-1}\frac{\upartial \mathbfss{A}}{\upartial \theta_{i}}\right)+\frac{1}{2}\mbox{trace}\left(\mathbfss{N}^{-1}\frac{\upartial \mathbfss{N}}{\upartial \theta_{i}}\right).\nonumber
\end{align}
The derivative computation provided in \autoref{eq-dLdt} is the general form for computing the gradient for any basis function definition, the only difference is the definition of $\frac{\upartial \Phi}{\upartial \theta}$. However, if computed naively, the computation can be time consuming since the partial derivatives will be mostly zeros for any given parameter in $\theta$ and some of the same operations are repeated. In the next section we provide a more efficient way to compute the gradient for RBF basis functions.

\subsection{Efficient Optimization}
\label{sec-efficient-optimization}
For the case of RBF basis functions, we can compute the partial derivatives more efficiently by first defining $\Delta_{j}=\mathbfss{X}-\mathbfit{1}_{n}\mathbfit{p}_{j}$, where $\mathbfit{1}_{n}$ is a vector of length $n$ consisting of all ones. We also first derive the partial derivatives with respect to $\bar{\mathbfit{w}}$, $\ln\bbeta$ and $\ln\Phi$:
\begin{align}
\frac{\upartial \bar{\mathbfit{w}}}{\upartial \ln\balpha} =& -\Sigma^{-1}\mathbfss{A}\bar{\mathbfit{w}},\\
\frac{\upartial \ln p\left(\mathbfit{y}\right)}{\upartial \ln\bbeta}=&-\frac{1}{2}\mathbfss{B}\bdelta^{2}-\frac{1}{2}\mathbfss{B}\left(\Phi\circ\left(\Phi\Sigma^{-1}\right)\right)\mathbfit{1}_{m}+\frac{1}{2},\\
\frac{\upartial \ln p\left(\mathbfit{y}\right)}{\upartial \ln\Phi}=&\left(\frac{\upartial \ln p\left(\mathbfit{y}\right)}{\upartial \ln\bbeta}\mathbfit{u}^{T}-\mathbfss{B}\bdelta\bar{\mathbfit{w}}^{T}-\mathbfss{B}\Phi\Sigma^{-1}\right)\circ \Phi,
\end{align}
where $\bdelta^{p}=\left\{\delta_{i}^{p}\right\}_{i=1}^{m}$ and similarly for other vectors. The symbol $\circ$ denotes the Hadamard product, i.e. element-wise matrix multiplication. The partial derivatives with respect to the parameters $\mathbfit{u}$, $b$, $\ln\balpha$ and $\ln\boldeta$ are as follows:
\begin{align}
\frac{\upartial \ln p\left(\mathbfit{y}\right)}{\upartial \mathbfit{u}}=&\Phi^{T}\frac{\upartial \ln p\left(\mathbfit{y}\right)}{\upartial \ln\bbeta}-\mathbfss{N}\mathbfit{u},\\
\frac{\upartial \ln p\left(\mathbfit{y}\right)}{\upartial b}=&\sum_{i=1}^{n}\frac{\upartial \ln p\left(\mathbfit{y}\right)}{\upartial \ln\beta_{i}},\\
\frac{\upartial \ln p\left(\mathbfit{y}\right)}{\upartial \ln\boldeta}=&-\frac{1}{2}\mathbfss{N}\mathbfit{u}^{2}+\frac{1}{2},\\
\frac{\upartial \ln p\left(\mathbfit{y}\right)}{\upartial \ln\balpha}=&-\left(\Phi^{T}\mathbfss{B}\bdelta\right)\circ\frac{\upartial \bar{\mathbfit{w}}}{\upartial \ln\balpha}-\frac{1}{2}\mathbfss{A}\bar{\mathbfit{w}}^{2}\\
&-\mathbfss{A}\bar{\mathbfit{w}}\circ\frac{\upartial \bar{\mathbfit{w}}}{\upartial \ln\balpha}-\frac{1}{2}\mbox{diag}\left(\mathbfss{A}\Sigma^{-1}\right)+\frac{1}{2}.\nonumber
\end{align}
The partial derivatives with respect to the parameters $\Gamma_{j}$ and $\mathbfit{p}_{j}$ of the pseudo points can be computed as follows:
\begin{align}
\frac{\upartial \ln p\left(\mathbfit{y}\right)}{\upartial \mathbfit{p}_{j}}= \frac{\upartial \ln p\left(\mathbfit{y}\right)}{\upartial \ln\Phi}\bigg[:,j\bigg]^{T}\Delta_{j}\Gamma_{j}^{T}\Gamma_{j},
\end{align}
\begin{align}
\frac{\upartial \ln p\left(\mathbfit{y}\right)}{\upartial \Gamma_{j}}= -\Gamma_{j}\left(\Delta_{j}\odot\frac{\upartial \ln p\left(\mathbfit{y}\right)}{\upartial \ln\Phi}\bigg[:,j\bigg]\right)^{T}\Delta_{j},\label{eq-gpvc-derivative}
\end{align}
where $\mathbfss{A}\odot \mathbfit{v}$ denotes a broadcast multiplication, i.e. an element-wise multiplication between the vector $\mathbfit{v}$ and each column vector in $\mathbfss{A}$. Note that if all basis are forced to share the same parameter $\Gamma$, then the partial derivative with respect to it is:
\begin{align}
\frac{\upartial \ln p\left(\mathbfit{y}\right)}{\upartial \Gamma}=\sum_{j=1}^{m}\frac{\upartial \ln p\left(\mathbfit{y}\right)}{\upartial \Gamma_{j}},\label{eq-gpgc-derivative}
\end{align}
we can also force $\Gamma_{j}$ to be a diagonal covariance, in which case the partial derivatives with respect to each $\mbox{diag}\left(\Gamma_{j}\right)$:
\begin{align}
\frac{\upartial \ln p\left(\mathbfit{y}\right)}{\upartial \mbox{diag}\left(\Gamma_{j}\right)}=\mbox{diag}\left(\frac{\upartial \ln p\left(\mathbfit{y}\right)}{\upartial \Gamma_{j}}\right),\label{eq-gpvd-derivative}
\end{align}
similarly, the basis functions can be forced to share a global diagonal $\mbox{diag}\left(\Gamma\right)$:
\begin{align}
\frac{\upartial \ln p\left(\mathbfit{y}\right)}{\upartial \mbox{diag}\left(\Gamma\right)}=\sum_{j=1}^{m}\mbox{diag}\left(\frac{\upartial \ln p\left(\mathbfit{y}\right)}{\upartial \Gamma_{j}}\right).\label{eq-gpgd-derivative}
\end{align}
In the case of variable length-scales, where $\Gamma_{j}$ is a scalar value $\gamma_{j}$, the partial derivative with respect to each $\gamma_{j}$ is:
\begin{align}
\frac{\upartial \ln p\left(\mathbfit{y}\right)}{\upartial \gamma_{j}}=\sum_{k=1}^{d}\frac{\upartial \ln p\left(\mathbfit{y}\right)}{\upartial \Gamma_{j}}\bigg[k,k\bigg],\label{eq-gpvl-derivative}
\end{align}
and to force all basis functions to have a global length-scale $\gamma$, the partial derivative is computed as follows:
\begin{align}
\frac{\upartial \ln p\left(\mathbfit{y}\right)}{\upartial \gamma}=\sum_{j=1}^{m}\sum_{k=1}^{d}\frac{\upartial \ln p\left(\mathbfit{y}\right)}{\upartial \Gamma_{j}}\bigg[k,k\bigg].\label{eq-gpgl-derivative}
\end{align}
The framework thus allows for six different configurations, variable full covariances ({\sc gpvc}) as in \autoref{eq-gpvc-derivative}, a global full covariance ({\sc gpgc}) as in \autoref{eq-gpgc-derivative}, variable diagonal covariances ({\sc gpvd}) as in \autoref{eq-gpvd-derivative}, a global diagonal covariance ({\sc gpgd}) as in \autoref{eq-gpgd-derivative}, variable scalar length-scales ({\sc gpvl}) as in \autoref{eq-gpvl-derivative} and a global scalar length-scale ({\sc gpgl}) as in \autoref{eq-gpgl-derivative}. The six configurations are all special cases of \autoref{eq-gpvc-derivative}; however, the computational cost can be greatly reduced by taking advantage of the simpler structures of the other configurations. 

\begin{figure}
        \centering
        \begin{subfigure}[b]{1\columnwidth}
                 \includegraphics[width=\textwidth]{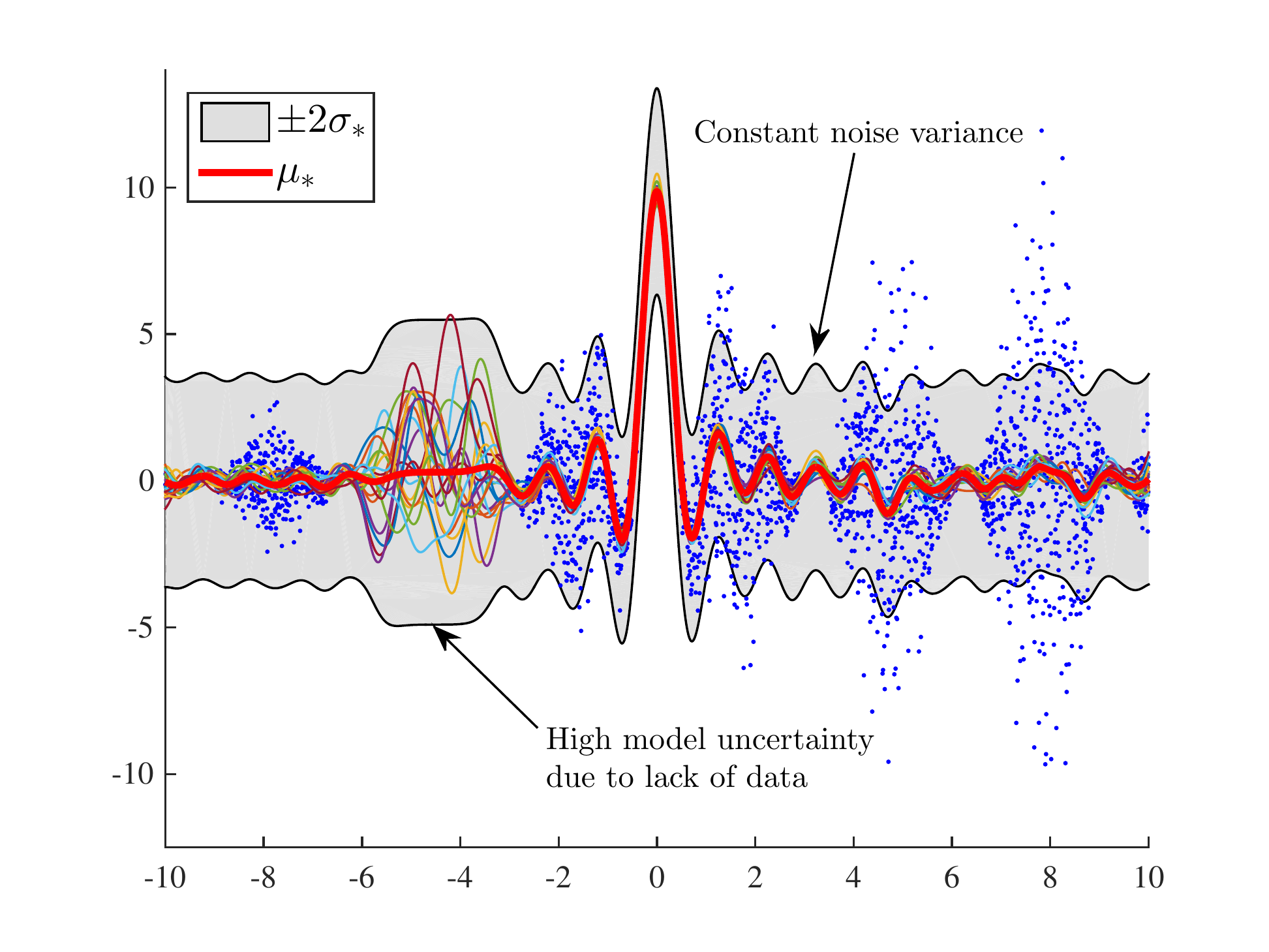}
                 \caption{Full GP}
        \end{subfigure}
        
        \begin{subfigure}[b]{1\columnwidth}
                 \includegraphics[width=\textwidth]{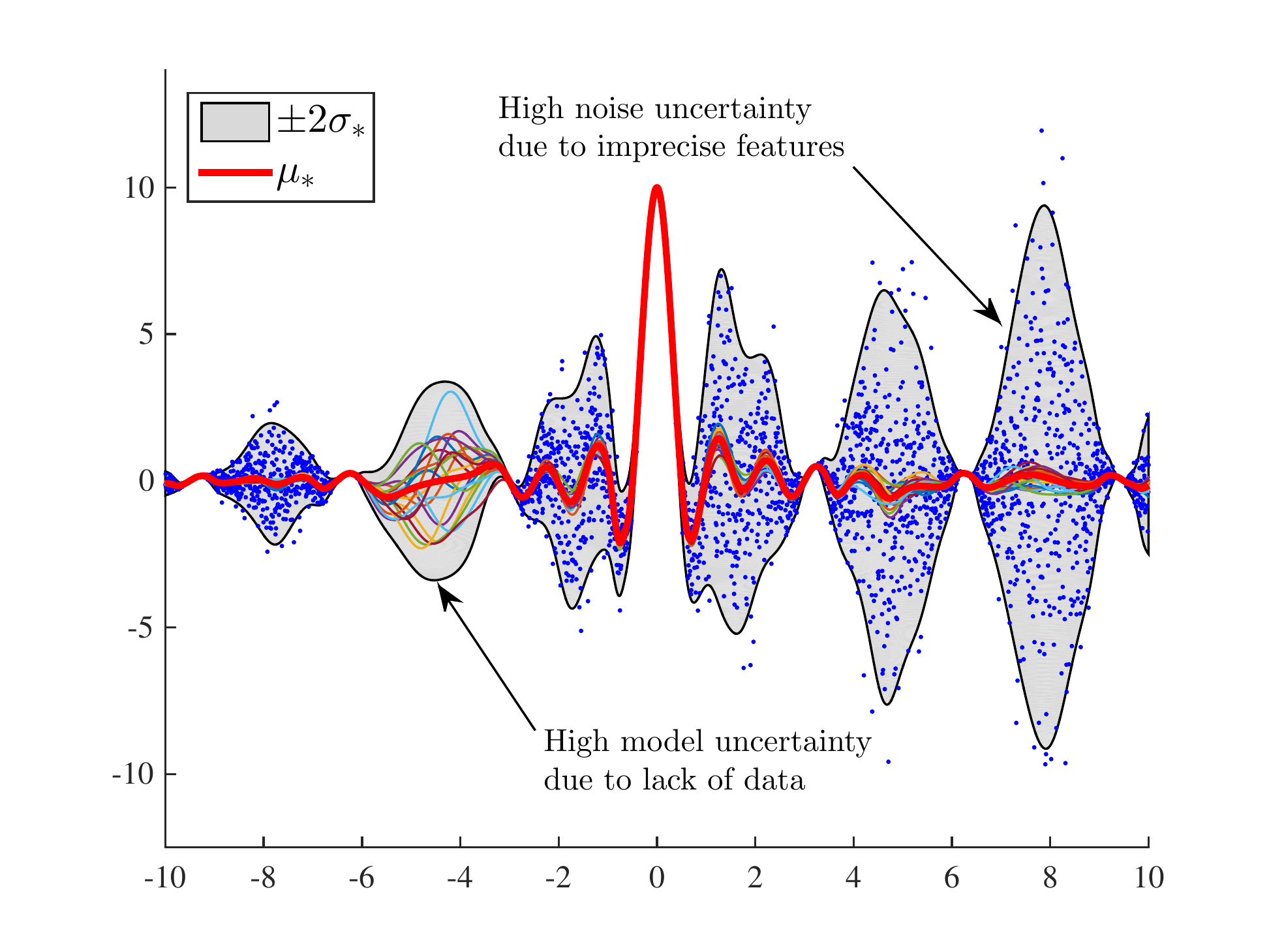}
                 \caption{{\sc gpvl}}
        \end{subfigure}
        
       \caption{The mean, variance ($\pm2\sigma_{*}$) and a sample of functions from the distribution produced by (a) a full GP model with a squared exponential kernel and (b) a {\sc gpvl} model trained using 200 basis functions. The generative distribution of the target output $y\left(x\right)\sim\mathcal{N}\left(\mu\left(x\right),\sigma^{2}\left(x\right)\right)$, where $\mu\left(x\right)=10\mbox{sinc}\left(2x\right)$ and $\sigma\left(x\right)=\frac{3\sin\left(x\right)}{1+\exp\left(-0.1x\right)}+0.01$ }
	\label{fig-sinc}
\end{figure}

In \autoref{fig-sinc}, we demonstrate the effect on a toy univariate example using a sparse GP with heteroscedastic noise and a full GP model with a squared exponential kernel. We used the {\sc gpml} toolbox implementation \citep{gpml2010} for the full GP model to offer a comparison to the variable length-scale basis function for the sparse GP ({\sc gpvl}). Note that both models estimate a higher predictive variance in the absence of data (-6 to -3 on the $x$-axis). However, this is the only source of uncertainty that the full GP is able to estimate accurately; the constraint of a constant noise variance has the negative effect of both over estimating and under estimating the true variance in different regions. On the other hand, the noise variance estimation in the {\sc gpvl} model is more accurate and leads to a more accurate determination of the total uncertainty about the mean function.

\begin{figure*}
        \centering
        
        \begin{subfigure}[b]{0.45\textwidth}
                 \includegraphics[width=\textwidth]{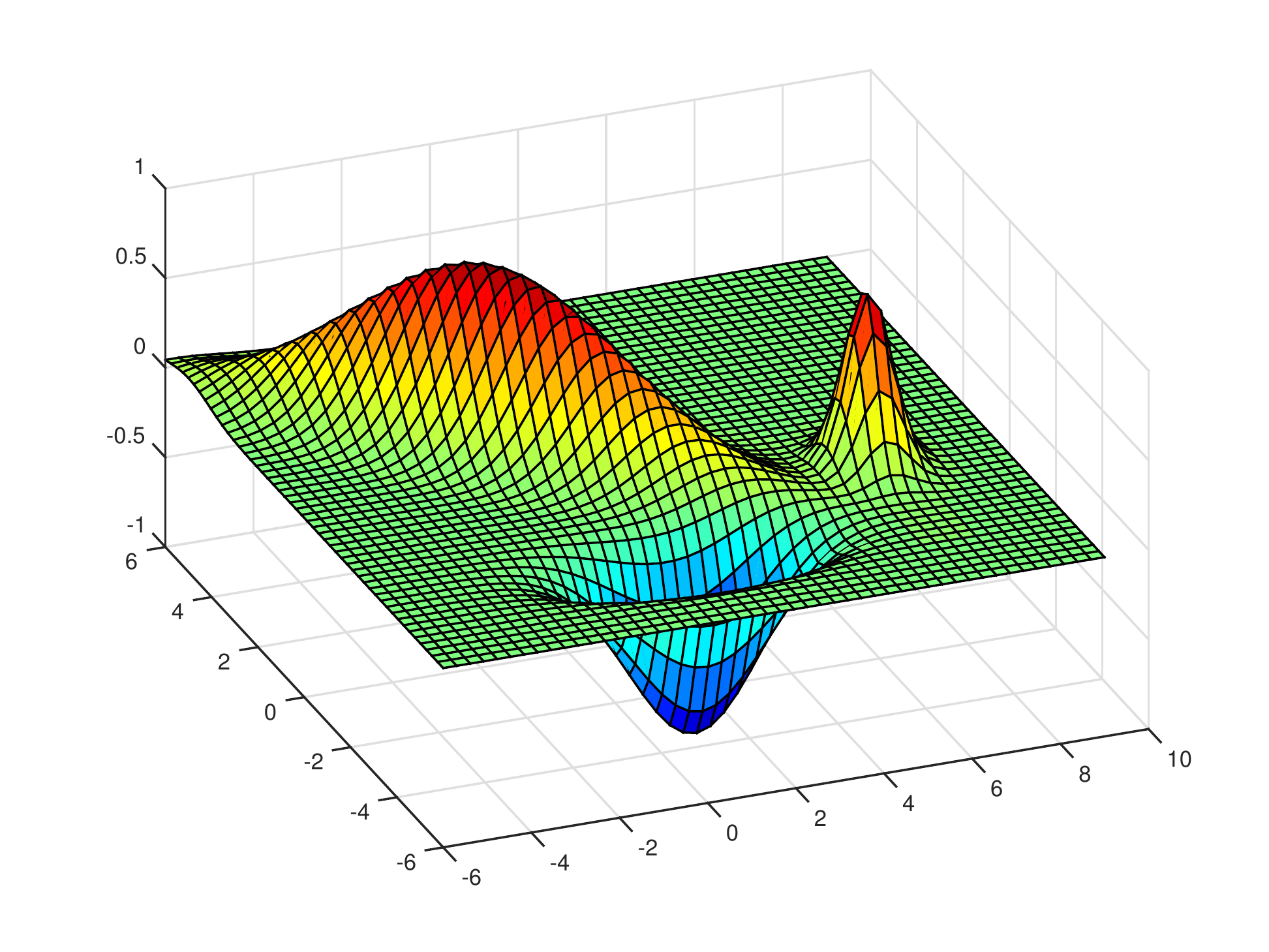}
                 \caption{Target function}
        \end{subfigure}
        ~
        \begin{subfigure}[b]{0.45\textwidth}
                 \includegraphics[width=\textwidth]{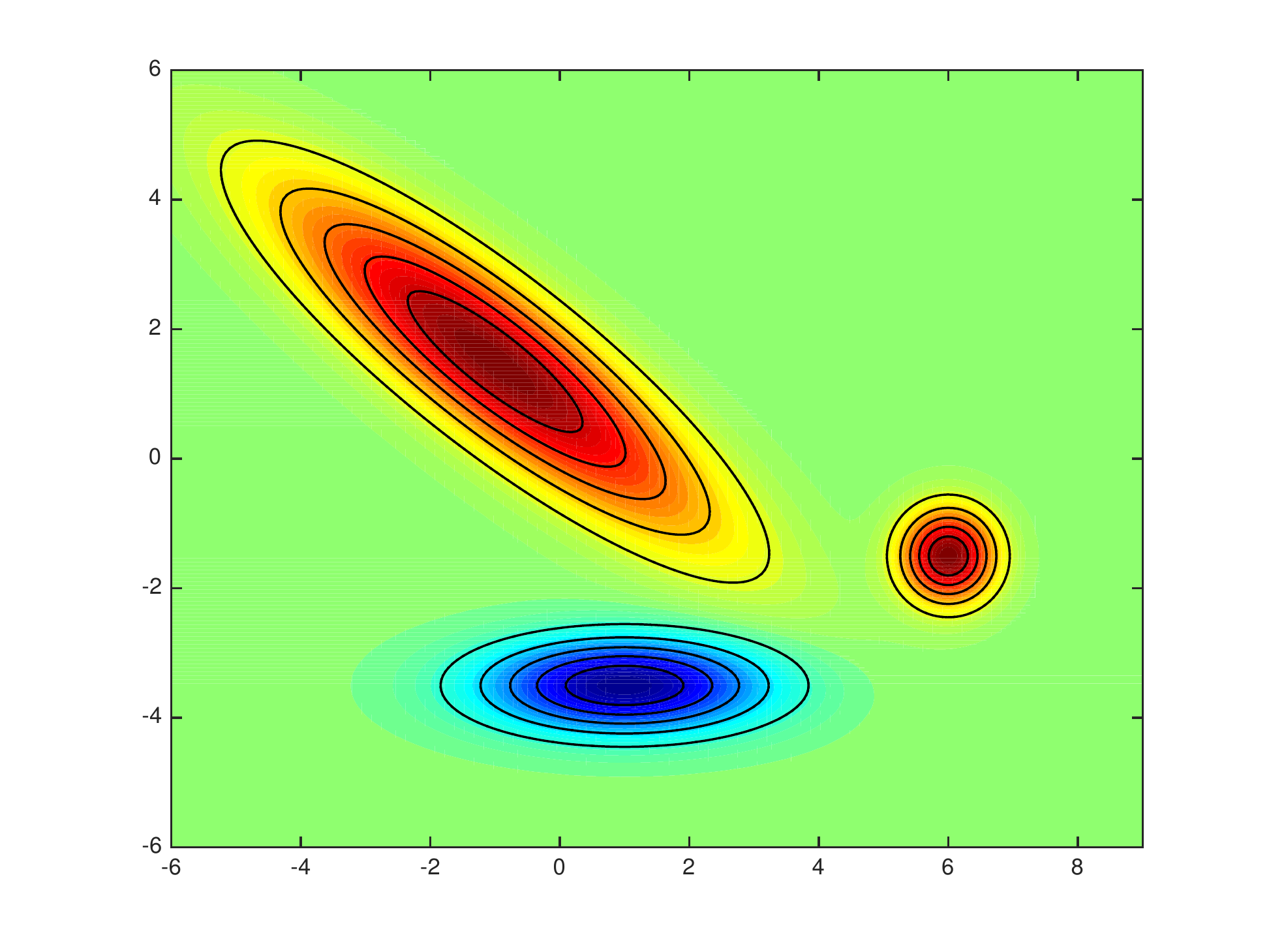}
                 \caption{Top view showing the true centres and covariances of the generating basis functions}
        \end{subfigure}
        
       \caption{A 2D synthetic example to illustrate the performance difference between {\sc gpvc}, {\sc gpgc}, {\sc gpvd}, {\sc gpgd}, {\sc gpvl} and {\sc gpgl}. The results are shown in \autoref{fig-2d-results}}
	\label{fig-2d-example}
\end{figure*}

\begin{figure*}
        \centering
        
        \begin{subfigure}[b]{0.3\textwidth}
                 \includegraphics[width=\textwidth]{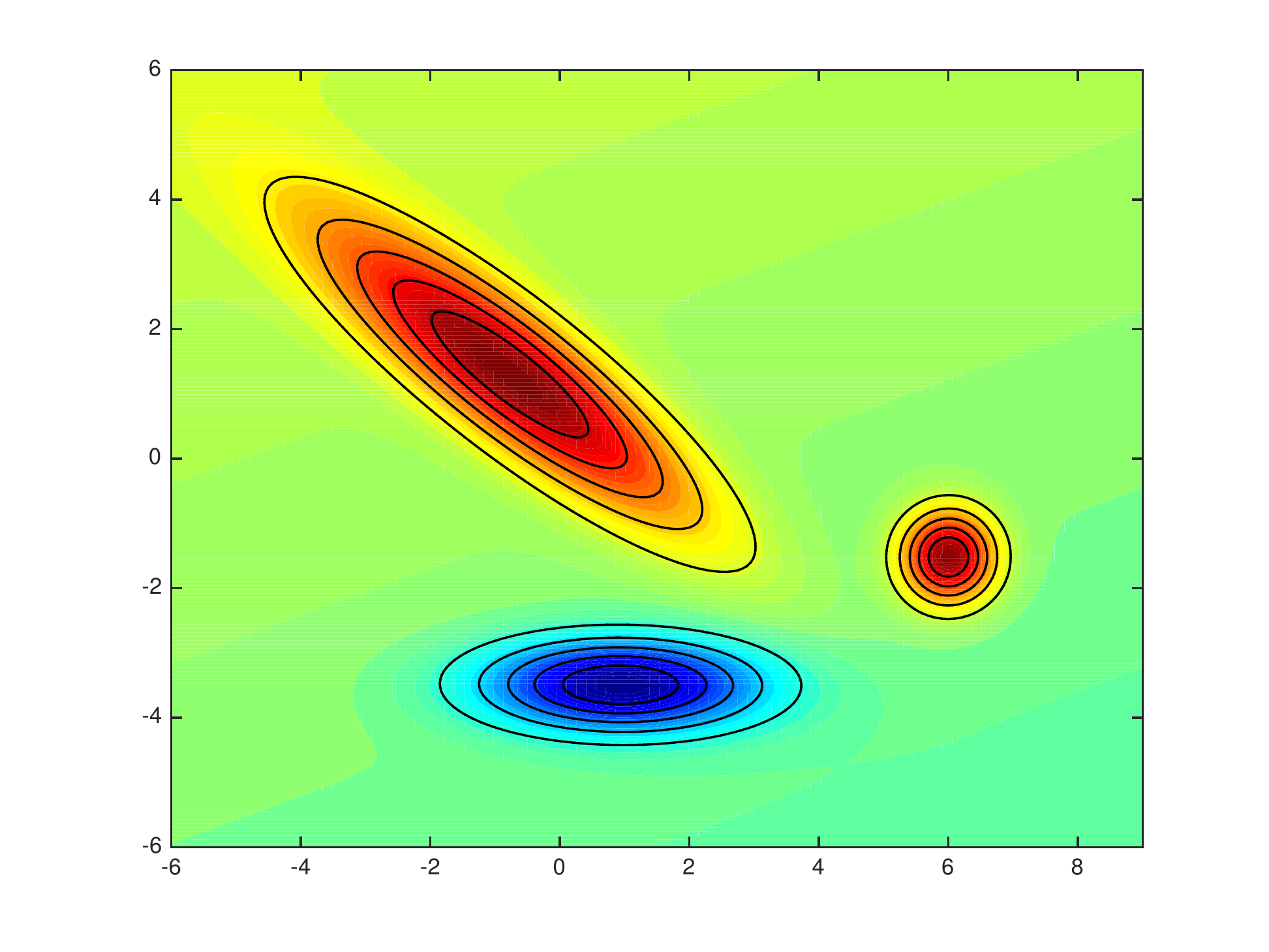}
                 \caption{{\sc gpvc} (RMSE=0.0589)}
        \end{subfigure}
        ~
        \begin{subfigure}[b]{0.3\textwidth}
                 \includegraphics[width=\textwidth]{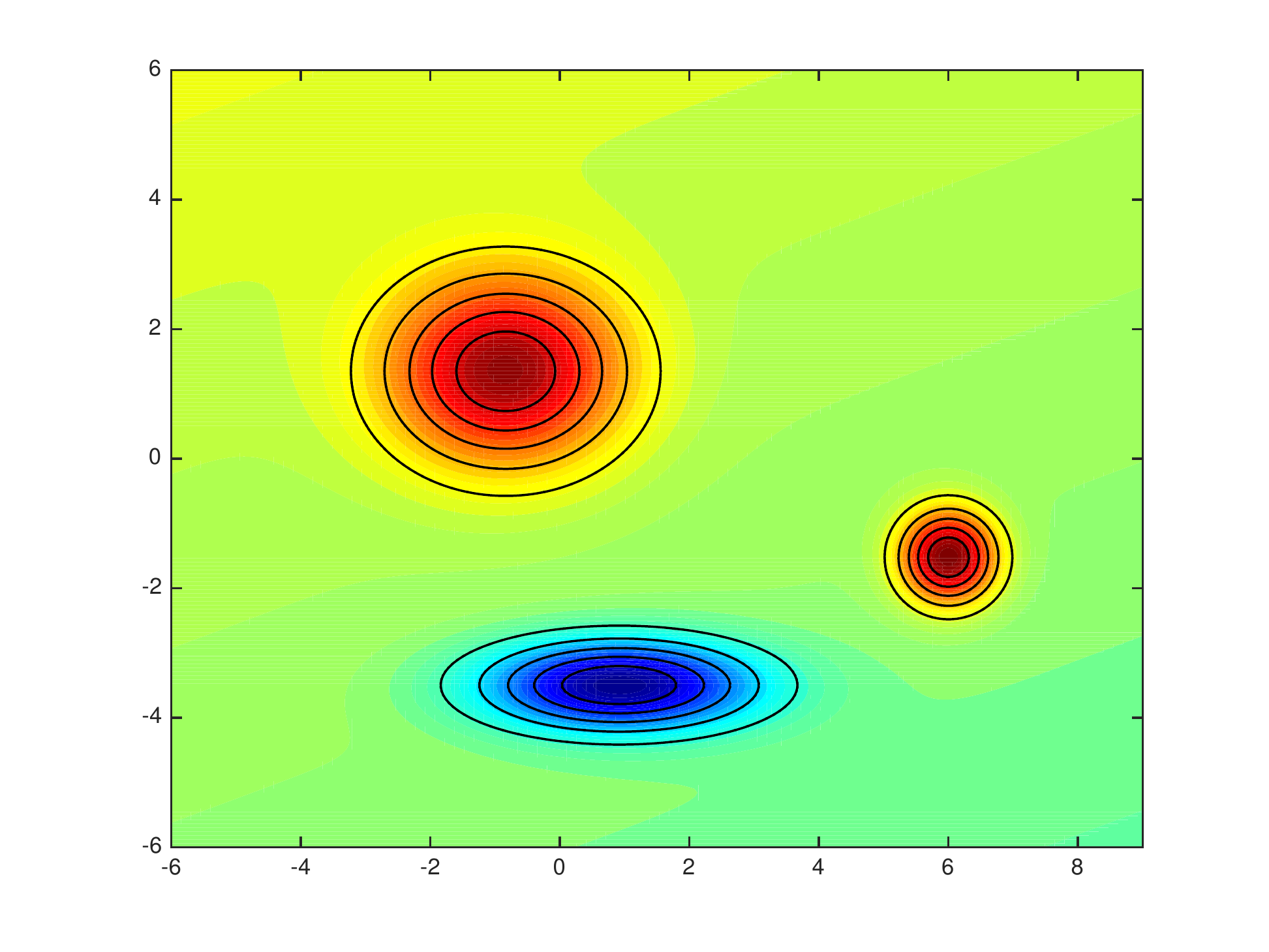}
                 \caption{{\sc gpvd} (RMSE=0.1140)}
        \end{subfigure}
        ~
        \begin{subfigure}[b]{0.3\textwidth}
                 \includegraphics[width=\textwidth]{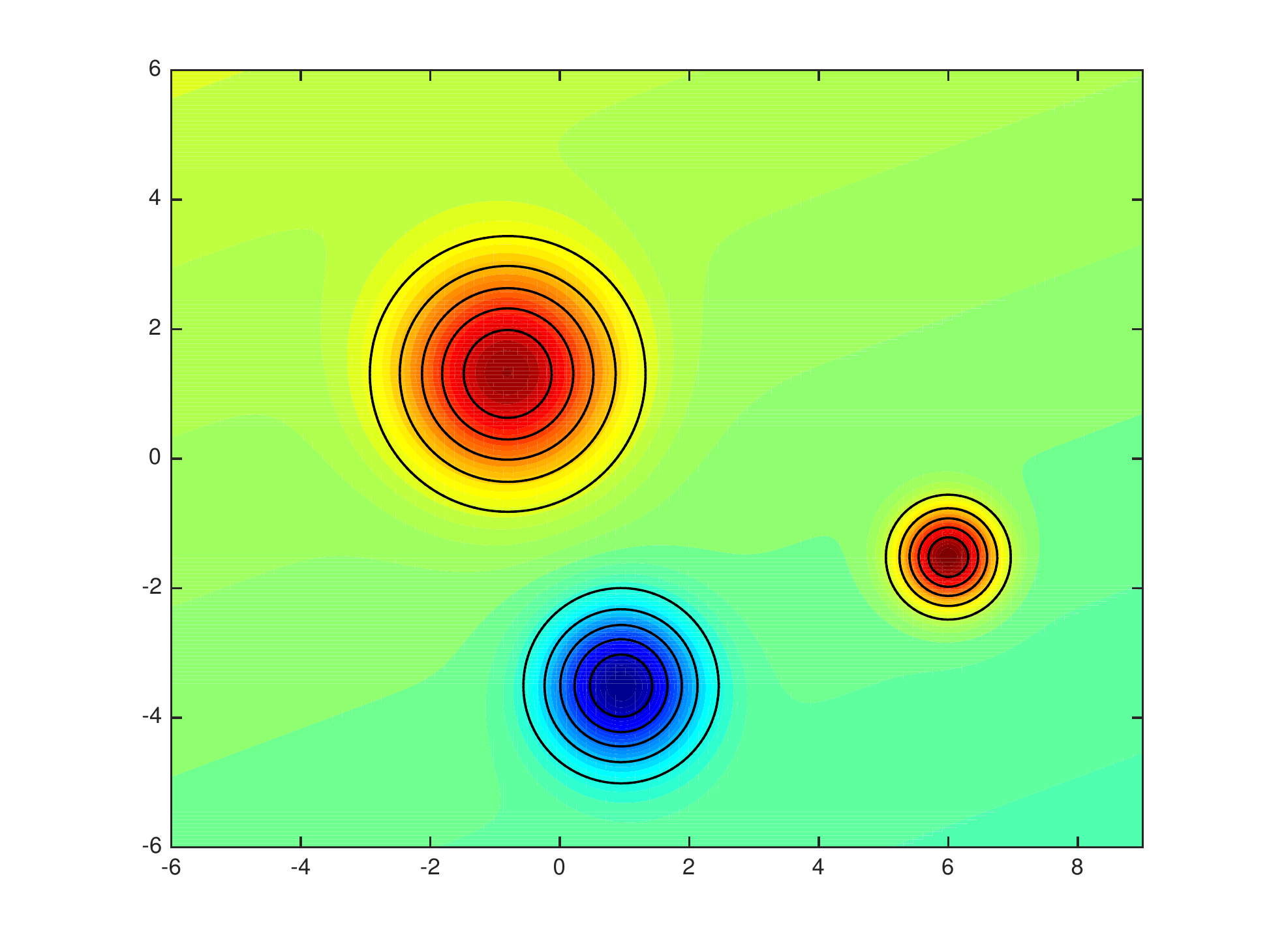}
                 \caption{{\sc gpvl} (RMSE=0.1305)}
        \end{subfigure}
         
        \begin{subfigure}[b]{0.3\textwidth}
                 \includegraphics[width=\textwidth]{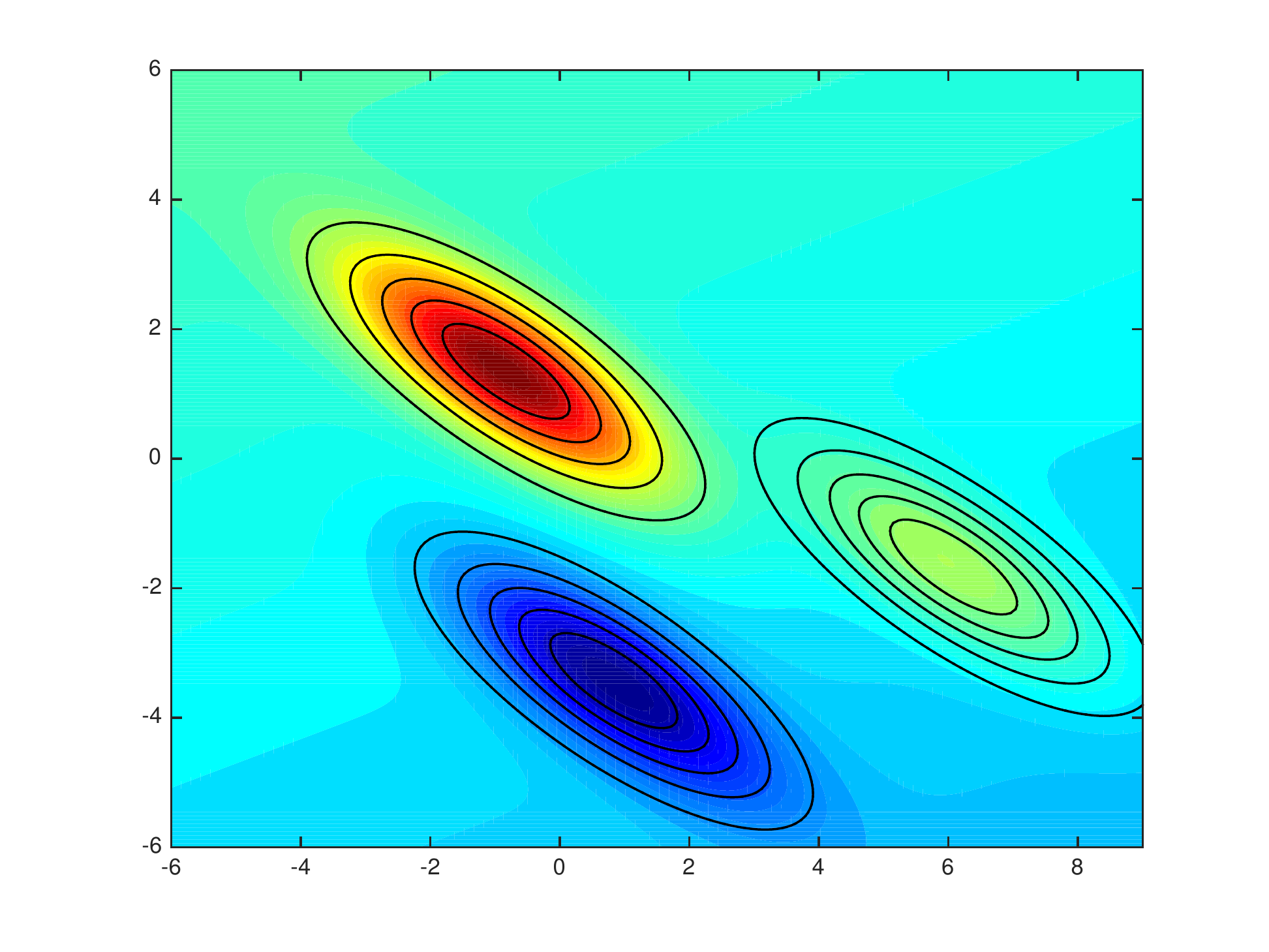}
                 \caption{{\sc gpgc} (RMSE=0.1077)}
        \end{subfigure}
         ~
        \begin{subfigure}[b]{0.3\textwidth}
                 \includegraphics[width=\textwidth]{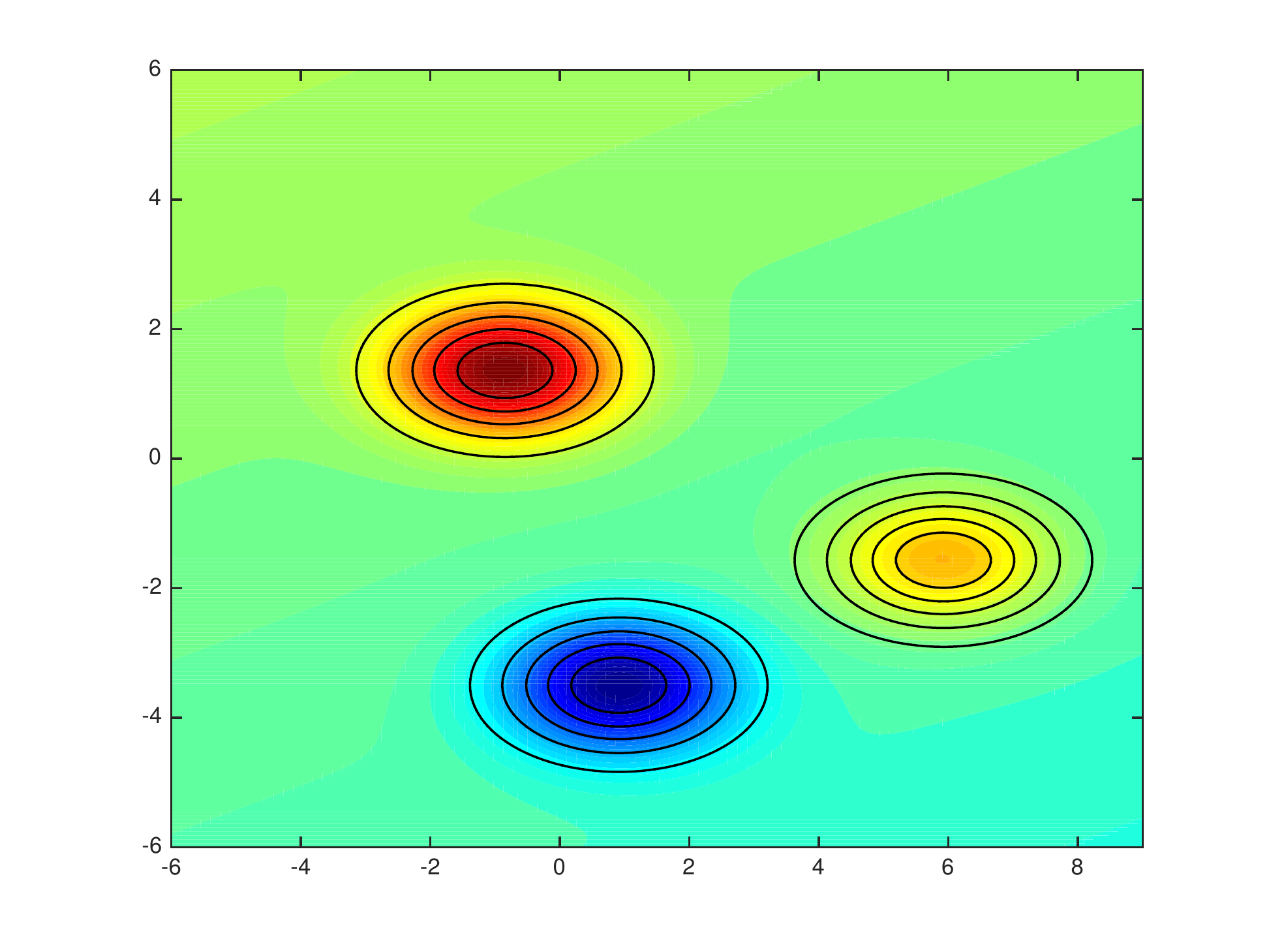}
                 \caption{{\sc gpgd} (RMSE=0.1290)}
        \end{subfigure}
         ~
        \begin{subfigure}[b]{0.3\textwidth}
                 \includegraphics[width=\textwidth]{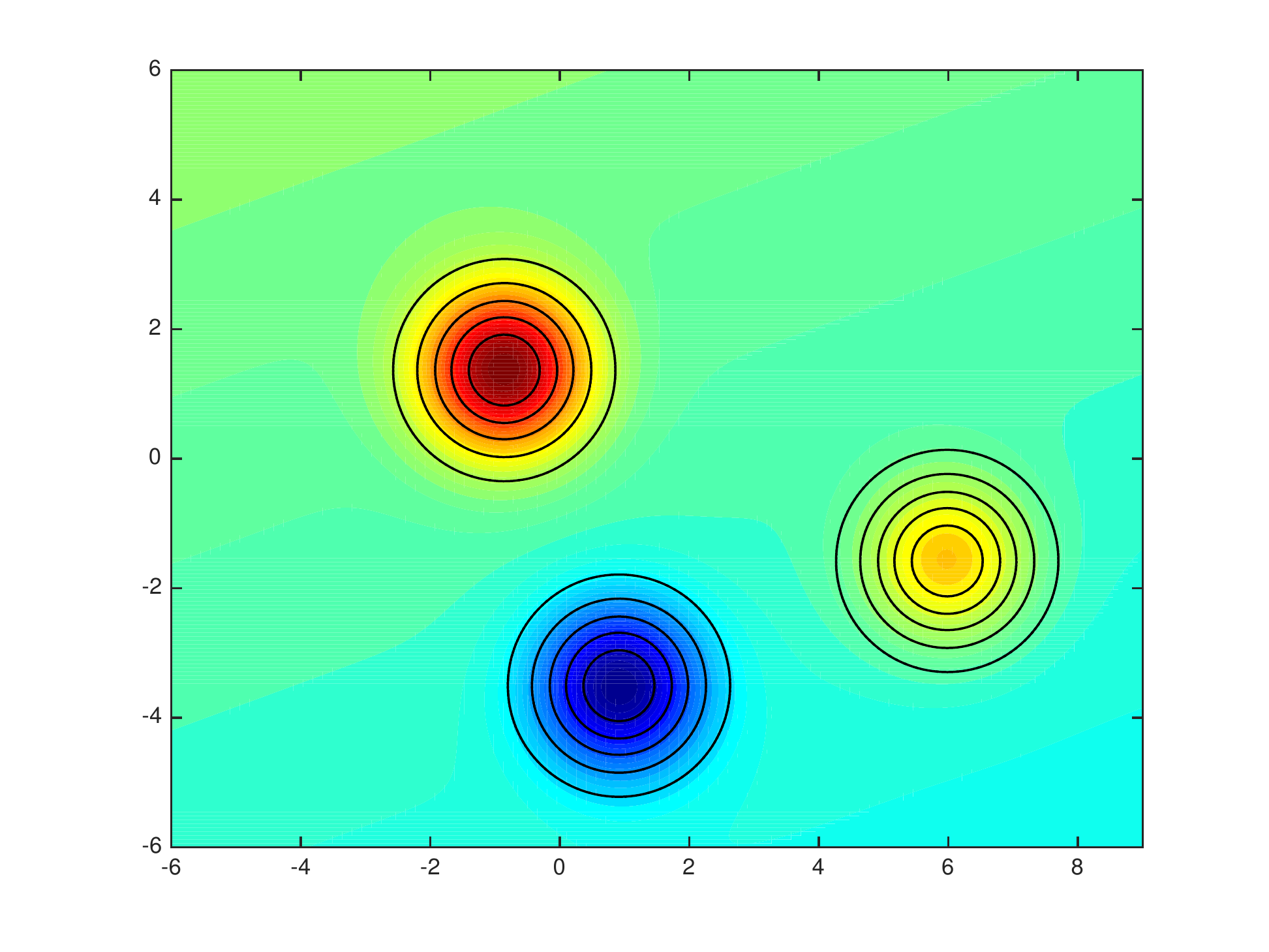}
                 \caption{{\sc gpgl} (RMSE=0.1363)}
        \end{subfigure}
        
       \caption{The results of running (a) {\sc gpvc}, (b) {\sc gpvd}, (c) {\sc gpvl}, (d) {\sc gpgc}, (e) {\sc gpgd} and (f) {\sc gpgl} using 3 basis functions on the synthetic 2D regression example in \autoref{fig-2d-example}. The RMSE performance on a held out test set for each are reported in subcaptions.}
	\label{fig-2d-results}
\end{figure*}

We illustrate the difference between the different configurations of the model, namely {\sc gpvc}, {\sc gpgc}, {\sc gpvd}, {\sc gpgd}, {\sc gpvl} and {\sc gpgl}, using a synthetic 2D example shown in \autoref{fig-2d-example}. The target function to be modelled is a linear combination of three basis function with different centres and covariances $f(x,y)=\phi_{1}(x,y)+\phi_{2}(x,y)-\phi_{3}(x,y)$. The different configurations were trained on examples of $x$ and $y$ as inputs and $f(x,y)$ as the target output plus some additive noise, the results are shown in \autoref{fig-2d-results}. It is not surprising that {\sc gpvc} performed the best, as it has more flexibility in modelling the covariance of each basis function. The other configurations would require more basis functions compared to {\sc gpvc} to achieve the same accuracy.

\section{SQL Statement}
\label{append-sql-statement}
The following SQL statement was used to extract the data from the SDSS DR12 database using the CasJobs service provided by SDSS\footnote{\url{casjobs.sdss.org}}.

\begin{verbatim}
SELECT
p.objid,
p.modelMag_u, p.modelMag_g,
p.modelMag_r, p.modelMag_i,
p.modelMag_z, p.modelMagerr_u,
p.modelMagerr_g, p.modelMagerr_r,
p.modelMagerr_i, p.modelMagerr_z,
s.z as zspec, s.zErr as zspecErr
INTO
mydb.modelmag_dataset
FROM
PhotoObjAll as p, SpecObj as s
WHERE
p.SpecObjID = s.SpecObjID AND
s.class = 'GALAXY' AND 
s.zWarning = 0 AND
p.mode = 1 AND
dbo.fPhotoFlags('PEAKCENTER') != 0 AND
dbo.fPhotoFlags('NOTCHECKED') != 0 AND
dbo.fPhotoFlags('DEBLEND_NOPEAK') != 0 AND
dbo.fPhotoFlags('PSF_FLUX_INTERP') != 0 AND
dbo.fPhotoFlags('BAD_COUNTS_ERROR') != 0 AND
dbo.fPhotoFlags('INTERP_CENTER') != 0
\end{verbatim}

\bsp	
\label{lastpage}
\end{document}